\documentclass[fleqn,usenatbib]{mnras}

\usepackage{newtxtext,newtxmath}

\usepackage[T1]{fontenc}

\usepackage{graphicx}	
\usepackage{amsmath}	
\usepackage{xcolor}
\usepackage{soul}
\usepackage{xspace}
\usepackage{longtable}
\usepackage{afterpage}
\usepackage{lineno}
\usepackage{hyperref}

\hypersetup{    
  colorlinks      = {true},
  linkcolor       = {blue},
  citecolor       = {blue},
  urlcolor        = {blue},
}

\newcommand{\code}[1]{\texttt{#1}\xspace}

\newcommand{\unit}[1]{\ensuremath{\mathrm{\,#1}}\xspace}
\newcommand{\feh}{\unit{[Fe/H]}}
\newcommand{\Teff}{\ensuremath{T_\mathrm{eff}}\xspace}
\newcommand{\logg}{\ensuremath{\log\,g}\xspace}
\newcommand{\vt}{\ensuremath{\nu_t}\xspace}

\newcommand{\nodata}{-}

\newcommand{\kms}{km~s\ensuremath{^{-1}}}

\newcommand{\msun}{\unit{M_{\odot}}}

\newcommand{\sfive}{\textit{S}$^5$}

\title[Multi-pop and CH Star in 300S Stellar Stream]{Multiple Populations and a CH Star Found in the 300S Globular Cluster Stellar Stream\thanks{This paper includes data gathered with the 6.5~meter Magellan Telescopes located at Las Campanas Observatory, Chile.}}
\author[S. A. Usman et al.]{
\parbox{\textwidth}{
\Large
Sam~A.~Usman,$^{1,2}$\thanks{Email: samanthausman@astro.uchicago.edu}
Alexander~P.~Ji,$^{1,2}$
Ting~S.~Li,$^{3,4,5}$
Andrew~B.~Pace,$^{6}$
Lara~R.~Cullinane,$^{7}$
Gary~S.~Da~Costa,$^{8,9}$
Sergey~E.~Koposov,$^{10,11,12}$
Geraint~F.~Lewis,$^{13}$
Daniel~B.~Zucker,$^{14,15}$
Vasily~Belokurov,$^{10}$
Joss~Bland-Hawthorn,$^{13,8}$
Peter~S.~Ferguson,$^{16}$
Terese~T.~Hansen,$^{17}$
Guilherme~Limberg,$^{1,2,18}$
Sarah~L.~Martell,$^{8, 19}$
Madeleine~McKenzie,$^{8, 20}$
and Joshua~D.~Simon$^{21}$
\begin{center} ($S^5$ Collaboration) \end{center}
}
\vspace{0.2cm}
\\
\parbox{\textwidth}{
$^{1}$ Department of Astronomy \& Astrophysics, University of Chicago, 5640 S Ellis Avenue, Chicago, IL 60637, USA\\
$^{2}$ Kavli Institute for Cosmological Physics, University of Chicago, Chicago, IL 60637, USA\\
$^{3}$ Department of Astronomy \& Astrophysics, University of Toronto, 50 St. George Street, Toronto ON, M5S 3H4, Canada\\
$^{4}$ Dunlap Institute for Astronomy \& Astrophysics, University of Toronto, 50 St George Street, Toronto, ON M5S 3H4, Canada\\
$^{5}$ Data Sciences Institute, University of Toronto, 17th Floor, Ontario Power Building, 700 University Ave, Toronto, ON M5G 1Z5, Canada\\
$^{6}$ McWilliams Center for Cosmology, Carnegie Mellon University, 5000 Forbes Ave, Pittsburgh, PA 15213, USA\\
$^{7}$ Department of Physics and Astronomy, Johns Hopkins University, 3400 N. Charles St, Baltimore, MD 21218, USA\\
$^{8}$ Centre of Excellence for All-Sky Astrophysics in Three Dimensions (ASTRO 3D), Australia\\
$^{9}$ Research School of Astronomy and Astrophysics, Australian National University, Canberra, ACT 2611, Australia\\
$^{10}$ Institute of Astronomy, University of Cambridge, Madingley Road, Cambridge CB3 0HA, UK\\
$^{11}$ Institute for Astronomy, University of Edinburgh, Royal Observatory, Blackford Hill, Edinburgh EH9 3HJ, UK\\
$^{12}$ Kavli Institute for Cosmology, University of Cambridge, Madingley Road, Cambridge CB3 0HA, UK\\
$^{13}$ Sydney Institute for Astronomy, School of Physics, A28, The University of Sydney, NSW 2006, Australia\\
$^{14}$ School of Mathematical and Physical Sciences, Macquarie University, Sydney, NSW 2109, Australia\\
$^{15}$ Macquarie University Research Centre for Astronomy, Astrophysics \& Astrophotonics, Sydney, NSW 2109, Australia\\
$^{16}$ Department of Physics, University of Wisconsin-Madison, Madison, WI 53706, USA \\
$^{17}$ Department of Astronomy, Stockholm University, AlbaNova University Center, SE-106 91 Stockholm, Sweden \\
$^{18}$ Universidade de S\~ao Paulo, IAG, Departamento de Astronomia, SP 05508-090, S\~ao Paulo, Brazil\\
$^{19}$ School of Physics, UNSW, Sydney, NSW 2052, Australia \\
$^{20}$ Research School of Astronomy \& Astrophysics, Australian National University, Canberra, ACT 2611, Australia\\
$^{21}$ Observatories of the Carnegie Institution for Science, 813 Santa Barbara St., Pasadena, CA 91101, USA
}
}
\date{Accepted XXX. Received YYY; in original form ZZZ}
\pubyear{2023}
\begin{document}
\label{firstpage}
\maketitle
%
\begin{abstract}
Milky Way globular clusters (GCs) display chemical enrichment in a phenomenon called multiple stellar populations (MSPs).
While the enrichment mechanism is not fully understood, there is a correlation between a cluster's mass and the fraction of enriched stars found therein. 
However, present-day GC masses are often smaller than their masses at the time of formation due to dynamical mass loss.
In this work, we explore the relationship between mass and MSPs using the stellar stream 300S.
We present the chemical abundances of eight red giant branch member stars in 300S with high-resolution spectroscopy from Magellan/MIKE.
We identify one enriched star characteristic of MSPs and no detectable metallicity dispersion, confirming that the progenitor of 300S was a globular cluster.
The fraction of enriched stars (12.5\%) observed in our 300S stars is less than the 50\% of stars found enriched in Milky Way GCs of comparable present-day mass ($\sim10^{4.5}$\msun).
We calculate the mass of 300S's progenitor and compare it to the initial masses of intact GCs, finding that 300S aligns well with the trend between the system mass at formation and enrichment.
300S's progenitor may straddle the critical mass threshold for the formation of MSPs and can therefore serve as a benchmark for the stellar enrichment process.
Additionally, we identify a CH star, with high abundances of \textit{s}-process elements, probably accreted from a binary companion.
The rarity of such binaries in intact GCs may imply stellar streams permit the survival of binaries that would otherwise be disrupted.
\end{abstract}




\section{Introduction}

\begin{table*}
    \caption{\label{tab:obs}The coordinates and exposure time for each star observed in this study. The distances were calculated using the relationship in \citet{Fu2018}. The Source column refers to the catalog or paper from which the star was chosen: \sfive~were stars observed by the \sfive~collaboration using AAT, Gaia stars were astrometrically selected from the Gaia catalog, and the \citet{Fu2018} star was the most metal-poor star in their sample.}
    \centering
    \setlength{\tabcolsep}{6pt}
    \scriptsize
    \begin{tabular}{cccccccccccc}
    \hline
\textbf{Gaia ID} & \textbf{RA} & \textbf{Dec} & \textbf{Star Name} & \textbf{G}$_0$ & \textbf{Distance} & \textbf{BP - RP} & \textbf{Radial Velocity} & \textbf{MJD} &\textbf{Exposure} & \textbf{Source} & \textbf{Notes} \\ 
        & ($^{\circ}$) & ($^{\circ}$) & & (mag, & (kpc) & (mag, & (\kms) & & \textbf{Time} (s) & &\\
        & & & & dereddened) &  & dereddened) & & & & \\ \hline
        3885023078597531520	 & 161.382227 & 14.149791 & J1045$+$1408 & 12.69 & 15.4 & 1.80 & 295.1 $\pm 0.8$ & 59342.0 &  600 & Gaia & Cool Star  \\
        &  &  &  &  &  &  & 297.0 $\pm$ 1.0 (GDR3) &  &  & \\ \hline
        3885197802162705152	& 162.428814 & 15.014678 &  J1049$+$1500 & 13.10 & 15.2 & 1.59 & 289.8 $\pm 0.8$ & 59342.0 &  900 & Gaia &\\
        &  &  &  &  &  &  & 291.1 $\pm$ 1.9 (GDR3) &  &  &  \\ \hline
        621582446059931264 & 151.605717 & 15.153214 & J1006$+$1509 & 13.99 & 17.4 & 1.50 & 307.1 $\pm 0.5$ & 59351.0 &  1200 &  \sfive &\\
        &  &  &  &  &  &  & 305.1 $\pm$ 0.7 (\sfive) & 58907.5 &  &  \\
        &  &  &  &  &  &  & 307.0 $\pm$ 3.9 (GDR3) &  &  &  \\ \hline
        622629352928532480 & 150.462821 & 15.900839 & J1001$+$1554 & 14.16 & 17.7 & 1.42 & 307.7 $\pm 1.0$ & 59351.0 &  1200 & \sfive & CH Star \\
         &  &  &  &  &  &  & 309.7 $\pm$ 0.8 (\sfive) & 58907.5 &  &  \\
        &  &  &  &  &  &  & 306.4 $\pm$ 2.6 (GDR3) &  &  &  \\ \hline
        622077466810856960 & 153.138229 & 15.908147 & J1012$+$1554 & 14.38 & 17.1 & 1.37 & 301.5 $\pm 0.8$ & 59342.0 &  1200 & \sfive & \\
        &  &  &  &  &  &  & 301.3 $\pm$ 0.7 (\sfive) & 59232.6 &  &  \\
        &  &  &  &  &  &  & 302.4 $\pm$ 3.6 (GDR3) &  &  &  \\ \hline
        621871205301503488 & 150.588462 & 15.691547 & J1002$+$1541 & 15.11 & 17.6 & 1.27 & 305.1 $\pm 0.5$ & 59342.0 &  2500 & \sfive & \\
        &  &  &  &  &  &  & 303.1 $\pm$ 0.9 (\sfive) & 58907.5 &  &  \\ \hline
        3888560349238428928 & 155.183933 & 15.925786 & J1020$+$1555 & 15.29 & 16.7 & 1.20 & 299.8 $\pm 0.7$ & 59342.0 &  5300 & \sfive & Enriched Star\\
        &  &  &  &  &  &  & 297.7 $\pm$ 1.0 (\sfive) & 58900.7 &  &  \\ \hline
        616868118157193344 & 149.225325 & 16.062778 & J0956$+$1603 & 15.65 & 17.9 & 1.18 & 304.3 $\pm 0.9$ & 59351.0 &  3000 & \sfive & \\
        &  &  &  &  &  &  & 302.8 $\pm$ 0.7 (\sfive) & 58908.5 &  &  \\ \hline
        \hline
        622146628669531008 & 153.057089 & 16.394690 & J1012$+$1623 & 17.80 & & & 261.5 $\pm 1.7$ (MIKE) & 59342 &  1800 & Fu18 & Non-member \\
         &  &  &  &  &  &  & 263.2 $\pm$ 2.1 (MIKE) & 59650.0 &  900 &  & Binary \\ 
         &  &  &  &  &  &  & 275.7 $\pm$ 2.2 (\sfive) & 59232.6 &  &  \\ 
         &  &  &  &  &  &  & 297.5 $\pm$ 6.0 (SEGUE) & 54174 &  &  \\ \hline
        3884319807767821312 & 157.492922 & 13.904273& J1029$+$1354 & 13.34 & & & 238.3 $\pm 0.5$ & 59342 &  900 & Gaia  & Non-member \\
        &  &  &  &  &  &  & 233.8 $\pm$ 1.9 (GDR3) &  &  &  \\
        \hline
    \end{tabular}
\end{table*}

The Milky Way's history of mergers with stellar systems can be traced through the remnants of dwarf galaxies and globular clusters (GCs) long ago disrupted by the galaxy's gravity \citep{Li2019, Li2022}.
These systems are tidally disrupted first into stellar streams, before eventually becoming fully mixed into the Milky Way halo.
For example, the remnants of GCs can be seen by nitrogen-enriched stars distributed throughout the outer reaches of the Galaxy \citep[e.g.,][]{Martell2011, Martell2016, Schiavon2017, Fern2020}.
Stellar streams from GCs can act as a clue to understand the dynamics and enrichment of their intact progenitors.
For example, outside of GCs, stars can be found to have unusual element enrichment due to accretion from a binary companion such as CH stars \citep[e.g.,][]{McClure1980, McClure1990,Luck1991,Lucatello2003,Hansen2016},
but GCs tend to disrupt binary systems well before mass transfer can occur due to their high density \citep{Cote1997}.
Spectroscopic observations of GCs or their resulting stellar streams could shed light on their stellar dynamics.

Perhaps the most interesting aspect of GCs is that most have multiple stellar populations (MSPs): groups of stars with correlated enrichment in elements such as N, Na, and Al and depletion in elements such as C, O and Mg \citep{Osborn1971, Bell1980, Cottrell1981, Norris1981}.\footnote{For the remainder of this paper, we refer to second-population stars as "enriched", despite being unusually \textit{depleted} in some elements.}
MSPs are thought to develop only during the formation of dense star clusters, though the exact mechanism is not well understood \citep[e.g.,][]{Carretta2006, Carretta2009, Bastian2018}.
The majority of globular clusters exhibit this unique correlated enrichment, though some unusual globular clusters like Ruprecht 106 have had no detected enriched stars \citep{Villanova2013}.
One clue to the nature of this phenomenon is the clear correlation between the mass of a globular cluster and its fraction of enriched stars \citep[e.g.,][]{Milone2017,Bastian2018,Gratton2019}.
The most massive intact Milky Way GCs contain higher fractions of enriched stars, with clusters of mass >$10^{4.5}$ \msun generally having at least half of their stars enriched \citep{Milone2017}.
The trend is not perfect, however: even the lowest-mass globular clusters can contain MSPs, contrary to expectations \citep{Simpson2017}.
The relationship between the fraction of enriched stars and the mass can be clarified when instead looking at the \textit{initial} mass of the systems instead of their present-day mas. 
\cite{Gratton2019} suggested that a globular cluster needs to achieve a minimum mass \emph{at its time of formation} in order to create MSPs.

Determining the initial mass of a globular cluster is not as straightforward as measuring the mass we see today, since the masses of GCs are not static \citep[e.g.,][]{Gnedin1997}.
Intact Milky Way GCs have typically lost 80\% of their initial mass due to tidal stripping, though the respective mass loss depends on each cluster's unique orbit \citep{Baumgardt2019}.
The critical mass for the formation of MSPs therefore appears lower when looking solely at current mass measurements of intact Milky Way GCs.
We can more closely examine the dependence of MSPs on initial mass by examining GCs accreted onto the Milky Way.
Additionally, comparisons of the fraction of enriched stars between Milky Way GCs and Magellanic Cloud GCs have found that the disruption of GCs via tidal forces may impact our understanding of mass and the enrichment of stars in GCs \citep{Milone2020}.
Stellar streams from disrupted GCs remain a useful and novel way to examine the relationship between the initial mass of a system and its enrichment.
Previous research on identifying MSPs in GCs have been performed on the GD-1 and Phoenix streams \citep{Wan2020, Casey2021, Balbinot2022}.
We turn to another ideal candidate for analysis, the 300S stellar stream.

The 300S stream was first found unintentionally by \citet{Geha2009} while studying the dwarf galaxy Segue 1, which overlaps 300S on the sky.
Its stars have since been re-observed spectroscopically by the Southern Stellar Stream Spectroscopic Survey, \sfive~\citep{Li2019, Li2022}.
From its highly eccentric and retrograde orbit, it is clear the stream's progenitor was accreted onto the Milky Way from another galaxy \citep{Fu2018}.
300S's progenitor may have formed in the galaxy Gaia-Sausage-Enceladus (\textit{GSE}, \citealt{Belokurov2018, Helmi2018}) prior to its merger with the Milky Way \citep{Fu2018, Li2022}.

In addition to its origin, the nature of the progenitor itself has also been debated over the past decade.
The stream has low abundances of $\alpha$ elements and has previously been found to have a metallicity dispersion, indicating that its progenitor was likely a dwarf galaxy \citep{Frebel2013, Fu2018}.
However, \citet{Li2022} used the Ca triplet to measure the metallicity, finding no detectable metallicity dispersion and concluding the progenitor of 300S was a globular cluster.

In this paper, we present a chemical abundance analysis of stars in 300S.
We show 300S is a globular cluster by identifying MSPs and finding no dispersion in metallicity.
Further, we detect a star enriched in carbon and \textit{s-}process elements, which we will refer to as a CH star.
In Section~\ref{sec:obs}, we describe the target selection and observations.
Section~\ref{sec:stellar_params} describes the determination of stellar parameters.
Section~\ref{sec:abund} describes the abundance analysis, including how the abundance dispersion was calculated.
The current stream mass and the initial mass of its progenitor are calculated in Section~\ref{sec:mass}.
The results are shown in Section~\ref{sec:results}.
We summarize our interpretations in Section~\ref{sec:discuss}.
Lastly, we discuss the broader implications for MSPs and future work in Section~\ref{sec:conclusion}.

\section{Observations}
\label{sec:obs}
We selected ten of the brightest candidate member stars of 300S in three ways (Table~\ref{tab:obs}).
First, we chose the six brightest 300S member stars found by \sfive~in \citet{Li2022}.
We refer readers to \citet{Li2022} for more details on the initial \sfive~data reduction and analysis, as well as the member star selection.
As \sfive~limited the targets to $G>14$ for 300S stream to avoid cross-talk between the brightest and faintest targets, we further selected three possible stars using Gaia DR3 \citep{GaiaCollaboration2016, GaiaCollaboration2022DR3} at a brighter magnitude (or beyond the \sfive~footprint) which could be potentially missed by \sfive, guided by the locations in the proper motion space and color-magnitude diagram of the previously identified 300S member stars from \citet{Li2022}.
Finally, we also included the most metal-poor star from \citet{Fu2018}, as this star is the main driver of the metallicity dispersion revealed by \citet{Fu2018}.

These ten stars were observed using the Magellan/MIKE high-resolution spectrograph on May 5th and 16th, 2021, using a 0.7$^{\prime \prime}$ slit and 2 × 2 on-chip binning \citep{Bernstein2003}.
We achieve a resolution of R $\sim$ 28,000 on the blue echelle orders and R $\sim$ 22,000 on the red echelle orders.
The data were reduced using \code{CarPy} \citep{Kelson2003}.
Details about the observations, including heliocentric radial velocities and exposure times, can be found in Table~\ref{tab:obs}.
The star J1020$+$1555 was also observed on April 13th, 2023 to increase the SNR, and the star J1012$+$1623 was observed on March 12th, 2022 to obtain an additional radial velocity.
We achieve SNRs measured per pixel ranging from 60 to 100 for red wavelengths around 8,000~\AA~and 35 to 50 for blue wavelengths around 4,500~\AA.

\begin{figure*}
    \centering
    \includegraphics[width=0.95\textwidth]{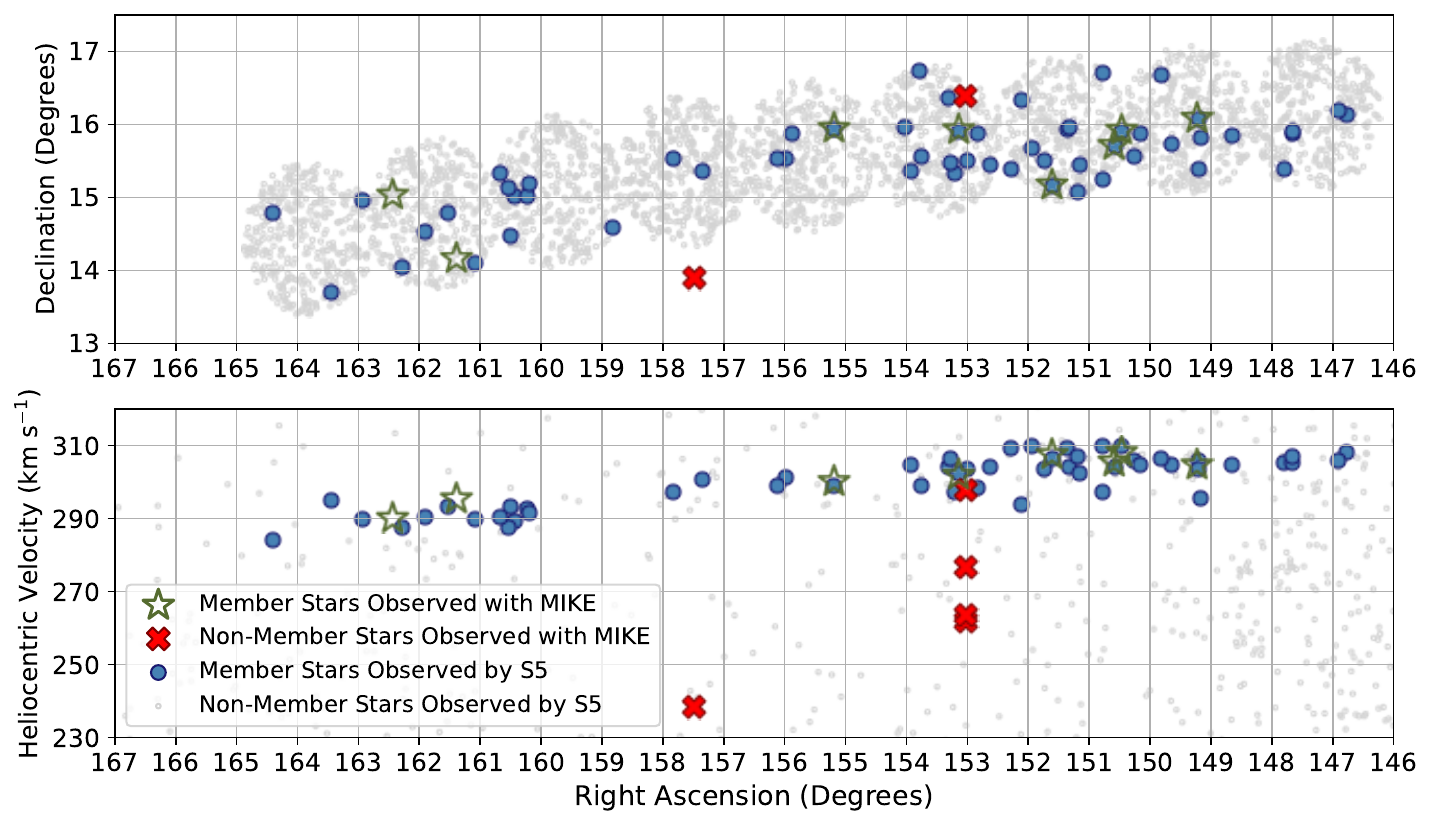}
    \caption{\label{fig:sky_map}The positions (top panel) and \sfive radial velocities (bottom panel) of stars in the region of 300S. Green stars represent 300S stars analyzed in this work. Red crosses represent stars rejected from this work as radial velocity non-members. The four red crosses to the right (at 153 degrees right ascension) represent the multiple radial velocities measured for the most metal-poor star analyzed by \citet{Fu2018}. Blue points represent 300S member stars identified in DECam photometry and \sfive~spectroscopy. Gray dots represent other stars observed by \sfive. }
\end{figure*}

Fig.~\ref{fig:sky_map} depicts the position and velocities of stars in the same region of sky as 300S.
We calculate the radial velocity by cross-correlating each echelle order with HD122563.
We take a weighted average of the orders' velocities to find the radial velocity of the star, down-weighting orders with high uncertainty.
The standard deviation of these velocities are taken to be the overall uncertainty of the star's radial velocity.
We also include radial velocities of stars obtained with \sfive \citep{Li2019} and Gaia DR3 \citep{Katz2023}.
Of the ten stars observed with MIKE, two were found to have radial velocities inconsistent with the stream track, indicating they are not members of 300S and were therefore rejected from this study.
In particular, there are four radial velocities for the most metal-poor candidate member star analyzed by \citet{Fu2018} (J1012$+$1623), showing clear velocity variations consistent with being in a close binary. 
We observed this star twice with MIKE to confirm its binarity. 
Though more velocities would be needed to robustly obtain a systemic velocity, the current data suggest that the SEGUE velocity was obtained at an unlucky phase of the binary orbit. Thus, we now classify this star as a radial velocity non-member.

\section{Stellar Parameter Determination}
\label{sec:stellar_params}

\begin{table*}\label{tab:stellar_params} 
\caption{
Stellar parameters.
[Fe/H] NLTE corrections were found using an online calculator based on \citet{Bergemann2012} and \citet{Lind2012}.
As this tool only covers parameter ranges 4000 < \Teff < 8000, 1.0 < \logg < 5.0, 1.0 < \vt < 2.0, for stars with parameters outside of these ranges we fix the respective inputs to \Teff = 4000 K, \logg = 1.0, and \vt = 2
}
\begin{tabular}{cccccccc}
\hline
\textbf{Star Name} & T$_{\text{eff}}$ (K) & \logg  & \vt \kms & [Fe/H] & $\Delta_{\text{[Fe/H] - [Fe/H]}_{\text{NLTE}}}$ & [Fe/H]$_{\text{NLTE}}$\\
\hline
J1045$+$1408 & 3880 & 0.20 & 2.39 & $-1.48$ & 0.004 & $-1.47$ \\
J1049$+$1500 & 4120 & 0.55 & 1.99 & $-1.38$ & 0.027 & $-1.35$ \\
J1006$+$1509 & 4230 & 0.85 & 2.06 & $-1.46$ & 0.036 & $-1.43$ \\
J1001$+$1554 & 4350 & 1.00 & 1.95 & $-1.38$ & 0.024 & $-1.36$ \\
J1012$+$1554 & 4420 & 1.13 & 2.06 & $-1.43$ & 0.048 & $-1.39$ \\
J1002$+$1541 & 4580 & 1.50 & 1.52 & $-1.27$ & 0.046 & $-1.22$ \\
J1020$+$1555 & 4710 & 1.68 & 1.86 & $-1.31$ & 0.045 & $-1.27$ \\
J0956$+$1603 & 4740 & 1.78 & 1.66 & $-1.32$ & 0.029 & $-1.29$ \\
\hline
\end{tabular}
\end{table*}

Four stellar parameters are needed to estimate stellar abundances: effective temperature T$_{\text{eff}}$, surface gravity log g, metallicity [Fe/H], and the microturbulence \vt.
The first two are determined using using Gaia DR3 data, which we dereddened using Equation~1 from \citet{Babusiaux2018} and reddening data from \citet{Schlegel1998} with applied corrections from \citet{Schlafly11}, specifically from the online database IRSA.\footnote{\url{https://irsa.ipac.caltech.edu/applications/DUST/}}
T$_{\text{eff}}$ is calculated using Equation~(1) from \citet{Mucciarelli2021}, using the coefficients for red giant branch stars:
\begin{align}
    \frac{5040}{T_{\text{eff}}} = 0.5472 + 0.5914 \times C +  0.2347 \times C^2 \nonumber \\
        -~0.0119 \times \text{[Fe/H]}  - 0.0012 \times \text{[Fe/H]}^2 \nonumber \\
        -~0.0012 \times \text{[Fe/H]} \times C \nonumber ~
\end{align}
Here, \textit{C} is the Gaia color (G-RP)$_0$ of the star.
We use [Fe/H] as measured spectroscopically for each star as given in Table~\ref{tab:stellar_params}.
We assume the star is in local thermodynamic equilibrium (LTE) and do not calculate non-LTE corrections (nLTE).
Most of our stars are in the range of 4000 K to 4800 K.
The most luminous star, J1045$+$1408, is slightly below this range at 3880 K.

Uncertainties in \Teff and \logg were estimated by propagating color and magnitude uncertainty.
The uncertainties for the \Teff equation from \citet{Mucciarelli2021} estimates an uncertainty of 71 K.
Each star had a unique color uncertainty based on the minimum and maximum estimates of reddening as calculated using dust maps from \citet{Schlegel1998} with corrections as applied by \citet{Schlafly11}.
We find these color uncertainties are 0.02 mag or less for all stars.
When propagating reddening uncertainties, the resulting \Teff error does not exceed 15 K.
For this reason, we conservatively round to a total uncertainty of 100 K for all of our stars.

The surface gravity \logg is calculated using Equation~3 from \citet{Venn2017}:
\begin{align}
    \log g = 4.44 + \log M + 4 \log\left(\frac{\Teff}{5780 \text{K}}\right) + \\
        0.4\times(g_0 - \mu + \text{BC}(M_g) - 4.75), \nonumber
\end{align}
where $\mu$ is the distance modulus, BC($M_g$) is the bolometric correction, and all constants are values for the Sun.
$M$ is the mass of the star, as interpolated from the best-fit isochrone.
The mass for all stars is found to be about 0.8 \msun, which is typical of old red giant stars.
We calculate $\mu$ from the distance gradient found by \citet{Fu2018}:
\begin{align}
    d~\text{(kpc)} = 48.9952-0.2083 \times \alpha
\end{align}
where $\alpha$ is the right ascension in degrees.
The bolometric corrections, BC, were interpolated from MARCS grids \citep{Casagrande2018} as a function of various stellar parameters.\footnote{\url{https://github.com/casaluca/bolometric-corrections/blob/master/bcutil.py}}
We used the maximum and minimum \Teff and reddening estimates to find the range of potential \logg values and used the extrema as estimates for our uncertainty.
Through this propagation of error, we found the uncertainty in \logg to be < 0.11 in all stars.
We propagate the uncertainty in distance to find an additional uncertainty of < 0.06 in all stars.
We adopt a total uncertainty of 0.2 for \logg.

The remaining stellar parameters used in the abundance analysis, microturbulence \vt and metallicity [Fe/H], were found using our spectroscopic measurements, as described in the next section.
Specifically, \vt is found by balancing line strength versus Fe I abundance, while [Fe/H] is manually set to match the spectroscopically measured [Fe I/H] metallicity.
The [Fe/H] error is the standard deviation found from the spectroscopic analysis.

\section{Abundance Analysis}
\label{sec:abund}

The spectra were analyzed using \code{SMHR}, a flexible interface for high-resolution spectral analysis \citep{Casey2014}.
We performed a standard analysis assuming local thermodynamic equilibrium (LTE) using the radiative transfer code \code{MOOG} \citep{Sneden1973, Sobeck2011} and 1D ATLAS stellar atmosphere models \citep{Castelli2004}.
We used \code{SMHR} to normalize and stitch together orders of the spectra, calculate and correct for radial velocities, fit equivalent widths, synthesize complex absorption regions, interpolate stellar atmospheres, and use curves-of-growth to determine abundances from absorption features.
To analyze the cool star, we used the further extended ATLAS catalog \citep{Meszaros2012}, which covers stellar atmospheres down to effective temperatures of 3500 K.
We adopted the line list used in \citet{Naidu2022}, which works well for red giants with [Fe/H] $\approx -1$. We used the line selection by \citet{Jonsson2017,Lomaeva2019,Forsberg2019} supplemented by \citet{Roederer2018}. The atomic data for the Fe, O, Mg, Ca, and Ti\,I lines was adopted from \citet{Jonsson2017}, and other elements were from \texttt{linemake}, \citep{Placco2021}.
For a more thorough description of how \texttt{SMHR} works, see \cite{Casey2014,Ji2020}.

\emph{Metallicity.}
Metallicity was determined from Fe I lines.
Our analysis in LTE returned an average metallicity of [Fe/H] = $-$1.39$\pm 0.03$.
The assumption of LTE is generally violated in metal-poor red giants, requiring non-LTE (NLTE) calculations.
We estimated corrections to the measured abundances to improve accuracy using calculations from \cite{Bergemann2019} using an online NLTE calculator.\footnote{\url{http://www.inspect-stars.com/cp/application.py/}}
This tool takes stellar parameters \Teff, \vt, and \logg (as well as \feh for non-Fe elements) and combines it with measured equivalent widths for specific lines to return corrections for each absorption line.
We calculate the correction for each line that is both measured in our analysis and available through the online calculator and apply it to our abundance measurements.
We then take the unweighted average for the corrected abundances and use this value as our NLTE-corrected measurements.
The calculator does not return uncertainties for the NLTE corrections, nor are we able to account for NLTE changes in our uncertainty measurements as our analysis pipeline assumes LTE.
We therefore do not include uncertainty for abundances after NLTE corrections have been applied.
We found the average NLTE-corrected abundance is [Fe/H] = $-$1.35, with a larger correction for cooler stars.

\emph{C, N, O.} Carbon abundances were inferred by fitting the spectral data with synthesized models in the range of 4300-4325 \AA. 
When available, nitrogen abundances were measured using synthesized models in the 3860-3890 \AA~CN~band.
These bands are too complex to measure with equivalent widths, unlike many other elements, including some of the iron-peak and neutron-capture elements.
O was measured using the forbidden line at 6300 \AA.
Telluric lines in the same region were fortunately moved when correcting to rest frame such that they are clearly identifiable at 6298/6299 and 6302/6303 \AA, far away from the forbidden O line.

\emph{$\alpha$-elements: Mg, Si, Ca.} Magnesium, silicon, and calcium were measured using equivalent widths in all eight stars.
Mg I was measured using four to seven absorption lines between 4500 and 5800 \AA.
Si I had up to six lines used, all between 5600 and 6000 \AA.
Ca I had up to seven lines in the range of 5800 and 6500 \AA.

\emph{Odd-Z Elements: Na, K.} Sodium and potassium were measured using equivalent widths.
Na was measured using two lines at 5682 and 5688 \AA.
K abundances were measured using equivalent widths of lines at 7699 and 7665 \AA.

\emph{Fe-Peak Elements: Ti, Mn, Ni, Zn.}
Titanium, manganese, nickel, and zinc were measured using a combination of synthesis and equivalent widths.
While Ti was not measurable in all stars, it could be measured in a few stars using equivalent widths of lines at 6091 and 6336~\AA.
Mn was measured using synthesis to allow for hyperfine splitting, primarily in the range of 4700-4900 \AA~and 6023 \AA.
Ni was mostly measured using equivalent widths between 6000 and 6400 \AA, but was also synthesized at 5476 \AA due to a blend.
Zn was measured at 4722 and 4810 \AA~using equivalent widths.

\emph{Neutron-Capture Elements: Sr, Y, Ba, Eu.}
Strontium, yttrium, barium, and europium were measured using a combination of synthesis and equivalent widths.
Sr and Eu were synthesized at 4215 \AA~and 6645 \AA, respectively.
Five lines were used to measure Ba at wavelengths from 4554 \AA~to 6496 \AA.
Since Ba has complex hyperfine and isotopic structure, we synthesize a range of absorption features for each line to properly measure the abundance \citep{Gallagher2012}.
For seven of the stars, we use an \textit{r}-process isotopic mixture, but for the CH star, J1001$+$1554, we use an \textit{s}-process isotopic mixture, as \textit{s}-process enrichment would be expected from accretion from a binary partner.
Four lines were used to measure Y at wavelengths from 4854 \AA~to 5200 \AA.

The measured abundances described here are shown in Table~\ref{tab:abundsummary}.

\subsection{Abundance Dispersion Analysis}
We calculate the intrinsic dispersion of all measured elements in the stars analyzed here.
We calculate the mean abundance and dispersion of the stream $\mu, \sigma$ given our measurements, $\vec{x}$ using Eqn.~\ref{eqn:likelihood}.
$\sigma^2_{\text{int}}$ and $\sigma_i^2$ represent the intrinsic dispersion and the individual error measurements, respectively, while $\mu$ is the mean abundance value.
The log likelihood is:
\begin{equation}
\label{eqn:likelihood}
\mathcal{L}(\mu, \sigma |\vec{x}) = \frac{-1}{2}\Sigma^N_{i=1}\left[\ln\left(2\pi(\sigma^2_{\text{int}}+\sigma_i^2)\right)+\frac{(x_i-\mu)^2}{(\sigma^2_{\text{int}}+\sigma^2_i)}\right]
\end{equation}
We then perform nested sampling over a range of combinations for the dispersion $\sigma^2_{\text{int}}$ and the mean $\mu$.
We use the nested sampling algorithm \code{dynesty} \citep{dynesty, Higson2019, Koposov2022} to calculate Bayesian posteriors for the cluster's abundance means and dispersions.
We include abundances and dispersions of the stream with and without anomalous stars to better demonstrate the significance of their exceptional abundances.
Results are given in Table~\ref{tab:disp}.
Most importantly, we find no significant dispersion in metallicity, indicating it is unlikely to be a dwarf galaxy.
A few elements such as Na and C have a measurable spread in abundances, mainly due to the anomalous stars such as the enriched star and the CH star, as will be discussed in Section~\ref{sec:results}.
We compare those stars to the main stellar population in Tables~\ref{tab:enrich} and \ref{tab:binary}, respectively.

\section{Stream Progenitor Mass Estimate}
\label{sec:mass}
We estimated the stellar mass of the 300S stream by comparing the number of confirmed red giant branch member stars to that of theoretical isochrones and their luminosity functions.
We did this for all eight stars analyzed in this work, as well as the remaining 34 300S member stars from \sfive \citep{Li2022}.
The \sfive~observations are sufficiently complete within its observation footprint to a \textit{Gaia} G magnitude of 17.2, as confirmed by matching the stream's observations with the theoretical luminosity function.

\begin{figure}
    \centering
    \includegraphics[width=0.5\textwidth]{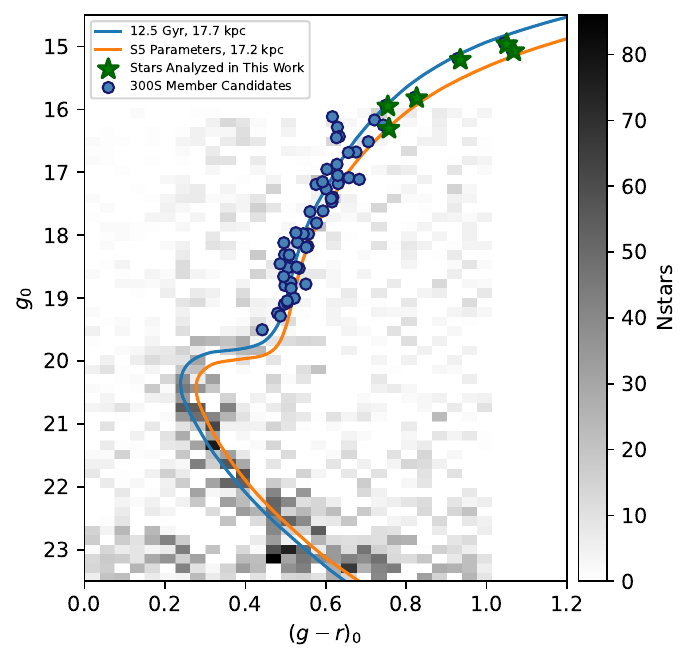}
    \caption{\label{fig:iso}Isochrones matched to DECam DECaLS DR9 photometry. The blue and orange lines represent the best-fit isochrones found in this article and using parameters from \citet{Fu2018} and \citet{Li2022}. The parameters used for the latter isochrone are: age = 12.5 Gyr, [$\alpha$/Fe] = 0.4, [Fe/H] $= -1.27$, d = 17.2 kpc, where [Fe/H] and d were from \citet{Li2022} and the age and [$\alpha$/Fe] abundance are obtained from \citet{Fu2018}. Our best-fitting isochrone parameters are: age = 12.5 Gyr, [$\alpha$/Fe] = 0.0, [Fe/H] $= -1.35$, d = 17.7 kpc. The green stars, and blue circles represent the dereddened color and magnitude of stars analyzed in this work and other member candidates observed by \sfive. The DECam data is a background-subtracted Hess diagram along 300S' sky position.}
\end{figure}

We used the Dartmouth Stellar Evolution Database \citep{Dotter2008}\footnote{\href{Isochrone Generation}{http://stellar.dartmouth.edu/models/}} to generate isochrones and visually fit them to our data.
The best-fit isochrone and the isochrone using stream parameters as measured by \citet{Fu2018} and \citet{Li2022} are shown in Fig.~\ref{fig:iso}.
We fit DECam DECaLs DR9 photometry for all 300S member stars observed by \sfive \citep{Dey2019}.
We created a background-subtracted Hess diagram to best match the isochrones to the main sequence turn-off.
From the isochrones, we find the cluster to be at a distance $d = 18.2 \pm 1$ kpc, have an age of 12.5 $\pm$ 1.5 Gyr, [Fe/H] = $-1.35^{+0.15}_{-0.05}$, and [$\alpha$/Fe] = 0.0.

We estimate that S5/AAT observations are complete for stars brighter than a Gaia \textit{G} magnitude of 17.2, as this is when our total number of observed stars falls below the predicted distribution from the luminosity functions.

We compare the proportion of stars we observe above our magnitude limit to the total mass of stars for the luminosity function.
We use a luminosity function using a Salpeter initial mass function \citep{Salpeter1955} and a 0.1 \msun mass cutoff.
This may likely be an overestimate for the current stream mass, as these low-mass stars may be preferentially stripped before others in the system due to mass segregation.
However, we are most interested in the stream's mass as a lower limit on the mass of the cluster at the time of formation.
The mass estimate with these assumptions may therefore more accurately reflect the mass of the system before tidal stripping occurred, though may be inflated compared to the present mass.

We then use the number of observed stars to solve for the total mass of the stellar stream in our system.
\begin{equation}
\label{eqn:mass_frac}
\frac{N_{\text{observable, lum func}}}{M_{\text{tot, lum func}}} = \frac{N_{\text{observable, 300S}}}{M_{\text{tot, 300S}}}
\end{equation}
Since we are observing red giant branch stars, the number of observable stars has stochastic error due to the short period of time the stars stay in this evolutionary stage.
We use the 1 $\sigma$ Poisson error as our upper and lower limits for the current mass for each isochrone.
The fixed magnitude detection limit means that the estimated mass depends on the distance, so we also use the highest and lowest distance estimates to estimate an error on the expected detection fraction.
We solve for the mass of the system with each isochrone with the predicted, upper and lower estimates for the mass, as well as our upper and lower distance measurements.
We choose the resulting extrema from these various combinations of parameters to use as limits on the mass of our system.

From this calculation, we find that the current mass of 300S is 10$^{4.6^{+0.3}_{-0.1}}$ \msun. 
However, we did not include stars which extend beyond \sfive's footprint for the stream. 
10$^{4.6}$ \msun is therefore a lower limit for the progenitor mass.
More sophisticated methods of calculating masses could be used in future investigations \citep[e.g.,][]{Martin2008, Martin2016}.  

We use a different method to estimate an upper limit on the stream mass:
Eqn.~5 from \citet{Baumgardt2003} gives the dissolution time $T_{\text{Diss}}$, the amount of time it would take for a globular cluster to completely disrupt given its initial mass $M_{\text{Ini}}$, initial number of stars $N_{\text{Ini}}$, and its orbital parameters assuming an isothermal mass distribution for the Milky Way:
\begin{equation}
\label{eqn:diss_time}
T_{\text{Diss}} = 1.35 \times \left(\frac{M_{\text{Ini}}}{\ln (0.02N_{\text{Ini}})} \right)^{0.75} \times \frac{R_{\text{Apo}}}{V_G} \times (1-\epsilon)
\end{equation}
Here, $R_{\text{Apo}}$, $V_G$, and $\epsilon$ are the distance at apocenter in kpc, the circular velocity of the Milky Way (assumed to be 240 \kms) and the eccentricity of the system's orbit, respectively.
Since 300S is completely dissolved and we can find no intact cluster, we use known orbital parameters and possible dissolution times to see the highest and lowest possible initial masses that would be disrupted by now.
We use the cluster's lower age limit and the age of the universe as minimum and maximum dissolution times respectively.
The cluster can disrupt any time after forming, so this is essentially a range for the upper limit of the dissolution time.
We used 300S's orbital parameters found in \citet{Li2022} and the best-fit age we found using isochrones.
We find the maximum initial mass of 300S to be M$_{\text{Ini}}$ = 10$^{4.8 \pm 0.1}$ \msun.

In conclusion, we find the progenitor mass of 300S is likely to be in the range of 10$^{4.5 - 4.9}$ \msun.
We note, however, that models of mass loss are under development and constraints on the initial mass may improve significantly with time.

\section{Results}
\label{sec:results}
We analyzed ten red giant branch member candidates of the 300S stream. 
We found two of the stars to be inconsistent with stream members (including the most metal-poor star analyzed in \citealt{Fu2018}), due to discrepant radial velocities. 
The eight analyzed stars had a mean LTE metallicity of [Fe/H] = $-1.38$ with an intrinsic dispersion of less than 0.09 at 95\% confidence level.
Abundances measured for each star can be found in Table~\ref{tab:abundsummary}.
Mean element abundances and dispersions are listed in Table~\ref{tab:disp}.
Of the eight stars, we highlight three in particular: the enriched J1020$+$1555, the CH star J1001$+$1554, and the cool star J1045$+$1408. \linebreak

\begin{figure*}
    \centering
    \includegraphics[width=\textwidth]{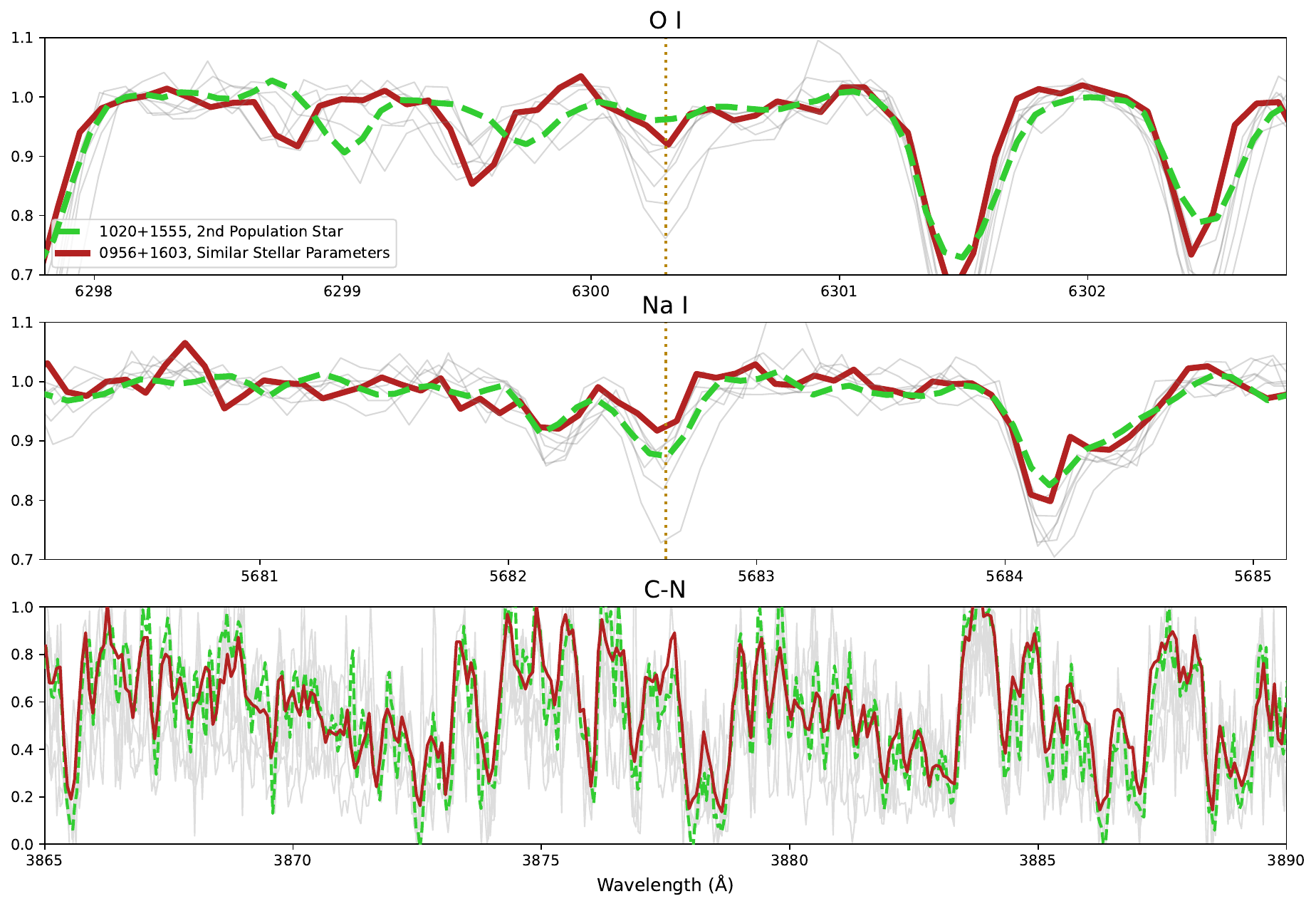}
    \caption{\label{fig:all_spec}A comparison of normalized spectra around an O line, two Na lines, and the C-N bands in the top, middle and lower panels, respectively. We highlight the Telluric lines in the top panel. We emphasize here the enriched star J1020$+$1555 (dashed green line) and the star with most similar stellar parameters, J0956$+$1603 (red solid line). The former has a higher Na and N enrichment and O depletion as expected of enriched stars in GCs. The spectra of the other stars, which are significantly colder than the highlighted stars, are shown as thin gray lines. The bright red dotted line indicates the wavelength of the measured absorption lines.}
\end{figure*}

\begin{figure*}
    \centering
    \includegraphics[width=\textwidth]{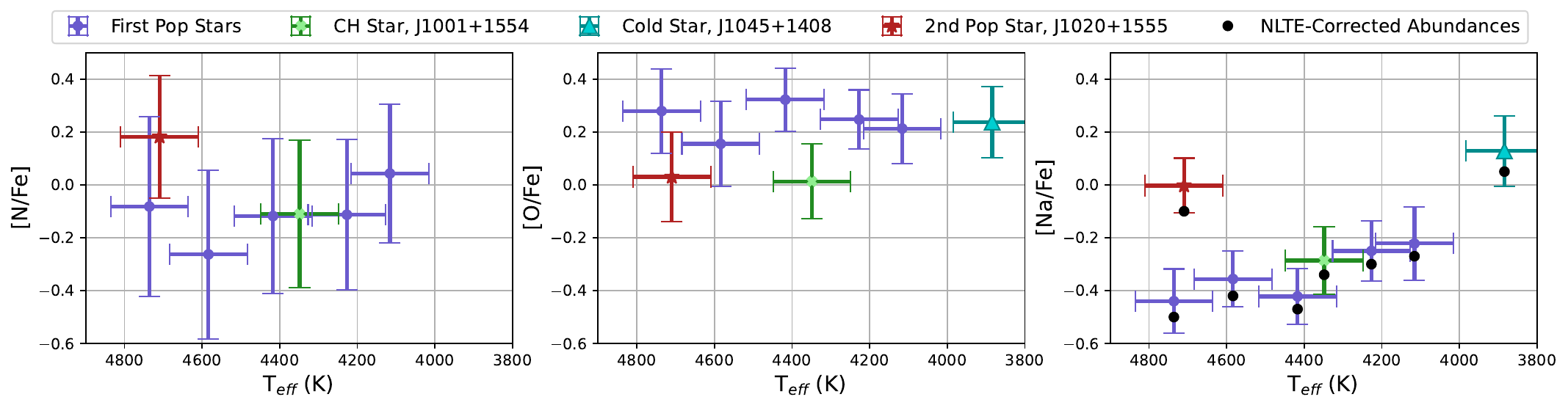}
    \caption{\label{fig:second_pop}The abundances of elements characteristic of multiple population enrichment: nitrogen, oxygen, and sodium in the left, middle and right plots, respectively. The purple circles represent the abundances of the ordinary stars. The anomalous stars J1020$+$1555, J1001$+$1554, and J1045$+$1408 are marked with a dark red star, a green cross, and a blue triangle respectively. J1020$+$1555 has the requisite enrichment in N and Na, and depletion in O. We also measure unusual of Na enrichment in the cool star, J0956$+$1603. Additionally, in the rightmost plot, the black points represent the NLTE-corrected abundances. However, due to the correlation between \Teff and the Na abundance and no depletion in O, we attribute this to an unknown systematic error and do not consider this a second-population star.}
\end{figure*}

\begin{figure*}
    \centering
    \includegraphics[width=0.9\textwidth]{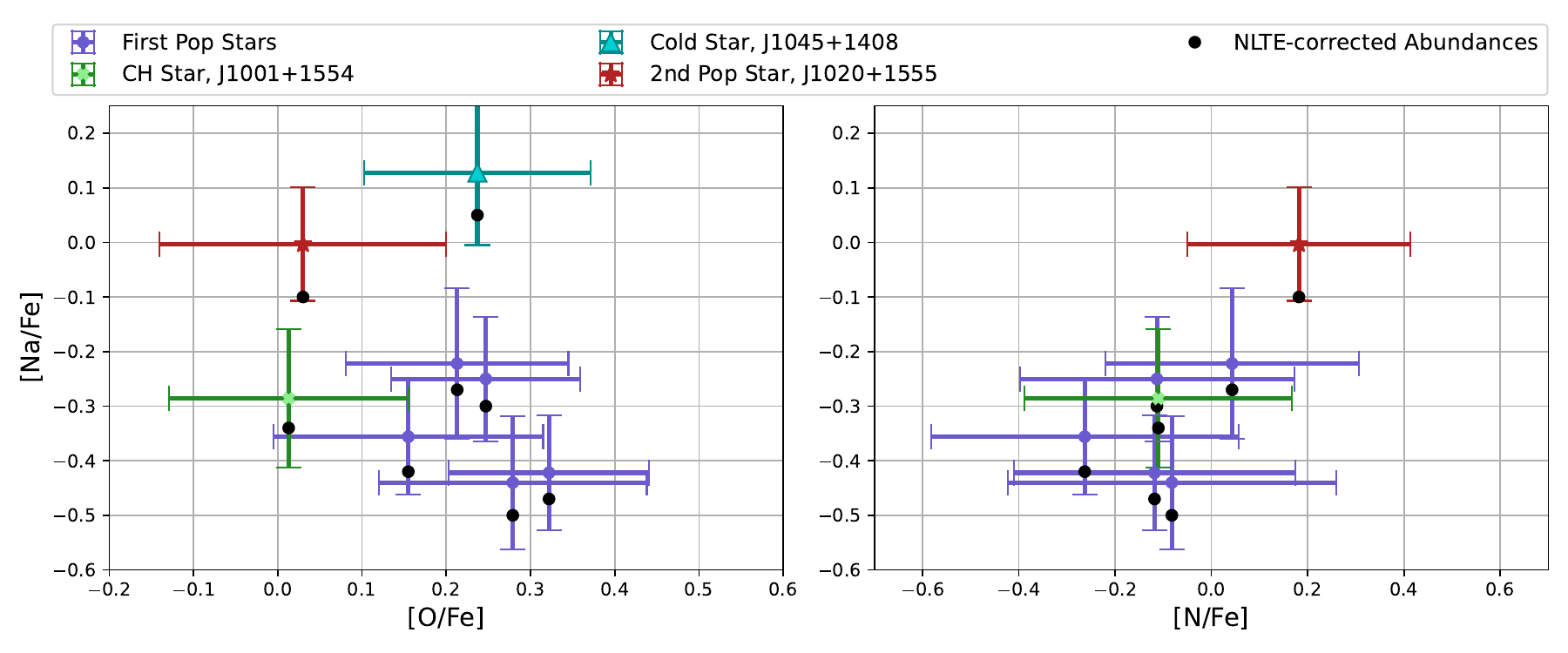}
    \caption{\label{fig:second_pop_corr}Correlated abundances of element Na compared to N and O in the left and right plots, respectively. The purple circles represent the abundances of the ordinary stars. The anomalous stars J1020$+$1555, J1001$+$1554, and J1045$+$1408 are marked with a dark red star, a green cross, and a blue triangle respectively. The black points represent the NLTE-corrected abundances. J1020$+$1555 has the requisite enrichment in N and Na, and depletion in O.  We also measure unusual of Na in the cool star, J0956$+$1603. However, due to the correlation between \Teff (as seen in Fig.~\ref{fig:second_pop}) and the Na abundance and no depletion in O, we attribute this to an unknown systematic error and do not consider this a second-population star.}
\end{figure*}

\emph{Enriched star: J1020$+$1555.} The second-warmest star of the group, J1020$+$1555, has the characteristic abundances of enriched stars found in GCs.
Specifically: J1020$+$1555 has enrichment in N and Na, and depletion in O.
We can see the differences in the absorption lines by comparing J1020$+$1555 to the 300S candidate with the most similar stellar parameters: J0956$+$1603.
In Fig.~\ref{fig:all_spec}, we can see that J1020$+$1555 (the dashed green line) has visually obvious differences in the absorption at key wavelengths compared to J0956$+$1603 (the solid red line).
The abundances as a function of \Teff in Fig.~\ref{fig:second_pop} and the abundance anti-correlations can be seen in Fig.~\ref{fig:second_pop_corr}.
Fig.~\ref{fig:second_pop} demonstrates that J1020$+$1555's anomalous abundances are prominent when compared to stars with similar \Teff.
The abundances are therefore not due to something like unusual stellar parameters, as could be the case in the cool star as we discuss later.
Fig.~\ref{fig:second_pop_corr} demonstrates that the abundances follow predicted correlations as observed in intact globular clusters.
Lastly we compare the data numerically and show the significance of the Na abundance scatter in Table~\ref{tab:enrich}.

\begin{figure*}
    \centering
    \includegraphics[width=0.9\textwidth]{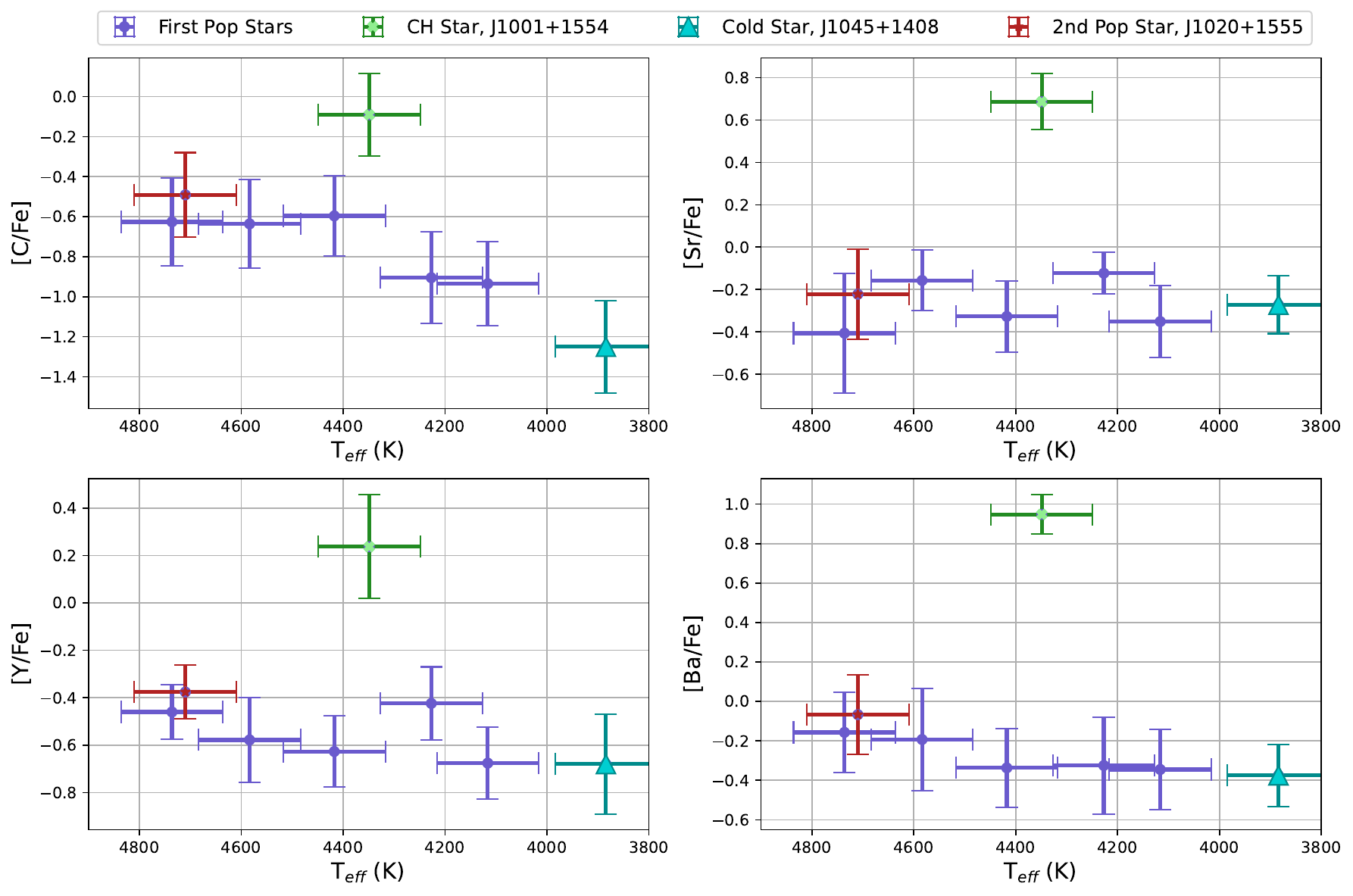}
    \caption{\label{fig:binary}The abundances of elements characteristic of a CH star: carbon and \textit{s}-process elements strontium, yttrium, and barium. The purple circles represent the abundances of the ordinary stars. The anomalous stars J1020$+$1555, J1001$+$1554, and J1045$+$1408 are marked with a dark red star, a green cross, and a blue triangle respectively. J1001$+$1554 has clear enrichment in all four of these elements.}
\end{figure*}

\emph{The CH star: J1001$+$1554.} This star has a significant amount of C and \emph{s}-process elements; specifically Sr, Y, and Ba have high abundances.
We can visually see the difference in the carbon and \textit{s}-process elements in Fig.~\ref{fig:binary} and numerically in Table~\ref{tab:binary}.
This enrichment indicates that J1001$+$1554 is a part of a binary star system, and its carbon and \emph{s}-process material were accreted from an AGB companion star \citep[e.g.,][]{McClure1990,Lucatello2003,Hansen2016}.
No significant radial velocity variations were detected from the three velocity measurements in Table~\ref{tab:obs}, though future more precise velocity measurements could potentially see a companion.
The enrichment found in this star is similar to that of CEMP-s stars or Ba stars \citep[e.g.,][]{Beers2005}; these stars may all have a similar enrichment mechanism involving mass transfer from a binary companion.
CH stars such as this one are uncommon in GCs, probably because the high density of the systems tend to disrupt binaries before they can reach the mass transfer phase \citep{Cote1997,DOrazi2010}.
We note that evolutionary mixing leads to changes in C abundances, which requires corrections to obtain the natal C abundance \citep{Placco2014}.
Though the abundances we present here do not include mixing corrections, the corrections would \textit{increase} C for the CH star, as C-enhanced stars have a larger corrections than stars with less carbon.

\emph{The cool star: J1045$+$1408.}
This star is the most luminous of the group, with an absolute Gaia 
\textit{G} magnitude of about $-$3.25 and effective temperature \Teff of 3880 K.
This star has a higher Na abundance than the other stars. 
There is a systematic correlating the Na abundance to \Teff, as can be seen in Fig.~\ref{fig:second_pop}.
Since this star is the coolest and has the lowest surface gravity of our stars, we expected larger NLTE effects than in our other stars.
We calculated NLTE corrections for the Na abundances for all stars using an online NLTE calculator based on \citet{Lind2011}.\footnote{\url{http://www.inspect-stars.com/cp/application.py/}}
The NLTE corrections to the Na abundances (derived using the 5682/5688{\AA} doublet) are about 0.05 to 0.10 dex and the resulting abundances can be seen in Fig.~\ref{fig:second_pop}.
The cool star's Na abundance still appears higher than those of the other stars, even when accounting for NLTE effects and the apparent \Teff systematic.
However, the cool star shows no O depletion as is found in the enriched star, and N was not measurable, so we have no further evidence it is an enriched star.
We conclude that the anomalous Na abundance is probably not a reflection of stellar enrichment, though we cannot rule it out.

\begin{table*}
\caption{The measured element abundance for each star for our measured elements. We were unable to measure N in the cool star, J1045$+$1408. We also include Na abundances which have had NLTE corrections been applied, though we are unable to calculate error for the corrected abundances as our analysis is fully LTE.}
\begin{tabular}{l|rr|rr|rr|rr|rr|rr|rr|rr|rr|rr|rrr}
\hline
\textbf{Star} & \multicolumn{2}{c}{\textbf{[Fe I/H]} $\sigma_{\text{[Fe I/H]}}$} & \multicolumn{2}{c}{\textbf{[Fe II/H]} $\sigma_{\text{[Fe II/H]}}$} & \multicolumn{2}{c}{\textbf{[C/Fe]} $\sigma_{\text{[C/Fe]}}$}& \multicolumn{2}{c}{\textbf{[N/Fe]} $\sigma_{\text{[N/Fe]}}$} & \multicolumn{2}{c}{\textbf{[O/Fe]} $\sigma_{\text{[O/Fe]}}$} & \multicolumn{3}{c}{\textbf{[Na/Fe]} $\sigma_{\text{[Na/Fe]}}$ ~\textbf{[Na/Fe]}$_{\text{NLTE}}$} \\
\hline
J1045$+$1408      &$-1.48$& $0.06$&$-1.23$& $0.15$&$-1.25$& $0.23$ & \nodata & \nodata &$+0.24$& $0.13$ & $+0.13$& $0.13$ & 0.05\\
J1049$+$1500      &$-1.38$& $0.06$&$-1.24$& $0.15$&$-0.94$& $0.21$ & $0.04$ & $0.26$ &$+0.21$& $0.13$ & $-0.22$& $0.14$ & $-0.27$\\
J1006$+$1509      &$-1.46$& $0.07$&$-1.45$& $0.14$&$-0.91$& $0.23$ & $-0.11$ & $0.29$ & $+0.25$& $0.11$ & $-0.25$& $0.11$ & $-0.30$\\
J1001$+$1554      &$-1.38$& $0.06$&$-1.44$& $0.10$&$-0.09$& $0.21$ & $-0.11 $ & $0.28$ & $+0.01$& $0.14$ & $-0.29$& $0.13$ & $-0.34$ \\
J1012$+$1554      &$-1.43$& $0.07$&$-1.31$& $0.09$&$-0.60$& $0.20$ & $-0.12$ & $0.29$ &$+0.32$& $0.12$ & $-0.42$& $0.11$ & $-0.47$ \\
J1002$+$1541      &$-1.27$& $0.09$&$-1.22$& $0.09$&$-0.64$& $0.22$ & $-0.27$ & $0.32$ &$+0.15$& $0.16$ & $-0.36$& $0.11$ & $-0.42$ \\
J1020$+$1555      &$-1.31$& $0.07$&$-1.38$& $0.11$&$-0.49$& $0.21$ & $0.18$ & $0.23$ &$+0.03$& $0.17$ & $-0.00$& $0.10$ & $-0.10$ \\
J0956$+$1603      &$-1.32$& $0.11$&$-1.34$& $0.12$&$-0.63$& $0.22$ & $-0.08$ & $0.34$ &$+0.28$& $0.16$ & $-0.44$& $0.12$ & $-0.50$ \\

\hline
& & & & & & & & & & & & \\
\hline

\textbf{Star} & \multicolumn{2}{c}{\textbf{[Mg/Fe]} $\sigma_{\text{[Mg/Fe]}}$} &  \multicolumn{2}{c}{\textbf{[Si/Fe]} $\sigma_{\text{[Si/Fe]}}$} &  \multicolumn{2}{c}{\textbf{[Ca/Fe]} $\sigma_{\text{[Ca/Fe]}}$} & \multicolumn{2}{c}{\textbf{[Sc/Fe]} $\sigma_{\text{[Sc/Fe]}}$} & \multicolumn{2}{c}{\textbf{[Ti/Fe]} $\sigma_{\text{[Ti/Fe]}}$} & \multicolumn{2}{c}{\textbf{[Mn/Fe]} $\sigma_{\text{[Mn/Fe]}}$} \\
\hline
J1045$+$1408    &$+0.16$& $0.11$ &$+0.40$& $0.08$ &$+0.04$& $0.12$ &$-0.53$& $0.14$ & $-0.20$& $0.19$ &$-0.51$& $0.18$ & \\
J1049$+$1500    &$+0.22$& $0.22$ &$+0.26$& $0.09$ &$+0.16$& $0.17$ &$-0.30$& $0.14$ & $+0.03$& $0.21$ &$-0.37$& $0.18$ & \\
J1006$+$1509    &$+0.11$& $0.11$ &$+0.29$& $0.09$ &$+0.16$& $0.15$ &$-0.19$& $0.13$ & $+0.11$& $0.19$ &$-0.42$& $0.14$ & \\
J1001$+$1554    &$+0.19$& $0.22$ &$+0.26$& $0.08$ &$+0.15$& $0.11$ &$-0.08$& $0.13$ & $+0.18$& $0.13$ &$-0.31$& $0.14$ & \\
J1012$+$1554    &$+0.04$& $0.11$ &$+0.24$& $0.07$ &$+0.18$& $0.09$ &$-0.21$& $0.13$ & $+0.12$& $0.15$ &$-0.40$& $0.14$ & \\
J1002$+$1541    &$-0.00$& $0.12$ &$+0.13$& $0.10$ &$+0.19$& $0.16$ &$-0.31$& $0.14$ & $-0.21$& $0.20$ &$-0.37$& $0.15$ & \\
J1020$+$1555    &$+0.07$& $0.09$ &$+0.17$& $0.08$ &$+0.13$& $0.15$ &$-0.10$& $0.12$ & $-0.18$& $0.16$ &$-0.33$& $0.16$ & \\
J0956$+$1603    &$-0.04$& $0.13$ &$+0.17$& $0.12$ &$+0.20$& $0.14$ &$-0.13$& $0.13$ & $+0.47$& $0.32$ &$-0.37$& $0.19$ & \\

\hline
& & & & & & & & & & & & \\
\hline

\textbf{Star} & \multicolumn{2}{c}{\textbf{[Ni/Fe]} $\sigma_{\text{[Ni/Fe]}}$} & \multicolumn{2}{c}{\textbf{[Zn/Fe]} $\sigma_{\text{[Zn/Fe]}}$} & \multicolumn{2}{c}{\textbf{[Sr/Fe]} $\sigma_{\text{[Sr/Fe]}}$} & \multicolumn{2}{c}{\textbf{[Y/Fe]} $\sigma_{\text{[Y/Fe]}}$} & \multicolumn{2}{c}{\textbf{[Ba/Fe]} $\sigma_{\text{[Ba/Fe]}}$} & \multicolumn{2}{c}{\textbf{[Eu/Fe]} $\sigma_{\text{[Eu/Fe]}}$} \\
\hline
J1045$+$1408  &$-0.14$& $0.24$ &$-0.09$& $0.17$ &$-0.27$& $0.14$ &$-0.68$& $0.21$ &$-0.38$& $0.16$&$+0.13$& $0.10$ & \\
J1049$+$1500  &$-0.13$& $0.11$ &$-0.41$& $0.11$ &$-0.35$& $0.17$ &$-0.68$& $0.15$ &$-0.34$& $0.20$&$+0.12$& $0.10$ & \\
J1006$+$1509  &$-0.06$& $0.11$ &$-0.25$& $0.12$ &$-0.12$& $0.10$ &$-0.42$& $0.15$ &$-0.33$& $0.25$&$+0.37$& $0.10$ & \\
J1001$+$1554  &$-0.23$& $0.18$ &$-0.41$& $0.11$ &$+0.69$& $0.13$ &$+0.24$& $0.22$ &$+0.98$& $0.11$&$+0.48$& $0.08$ & \\
J1012$+$1554  &$-0.08$& $0.13$ &$-0.20$& $0.11$ &$-0.33$& $0.17$ &$-0.63$& $0.15$ &$-0.34$& $0.20$&$+0.31$& $0.09$ & \\
J1002$+$1541  &$-0.03$& $0.15$ &$+0.01$& $0.16$ &$-0.16$& $0.14$ &$-0.58$& $0.18$ &$-0.19$& $0.26$&$+0.12$& $0.14$ & \\
J1020$+$1555  &$-0.12$& $0.14$ &$-0.15$& $0.13$ &$-0.22$& $0.21$ &$-0.38$& $0.11$ &$-0.07$& $0.20$&$+0.47$& $0.08$ & \\
J0956$+$1603  &$-0.17$& $0.22$ &$-0.25$& $0.15$ &$-0.41$& $0.28$ &$-0.46$& $0.12$ &$-0.16$& $0.20$&$+0.14$& $0.16$ & \\
\hline
\label{tab:abundsummary}
\end{tabular}
\end{table*}

\renewcommand{\arraystretch}{1.5}

\begin{table}
\caption{\label{tab:disp}The most probable mean abundance and dispersion of 300S as inferred from the element abundances measured in all eight stars. The error bars and upper limits shown here represent one standard deviation confidence. We constrain the metallicity dispersion to less than < 0.09 with 95\% confidence. Since our analysis was performed with the assumption of local thermodynamic equilibrium, we are unable to estimate the error for NLTE-corrected Na abundances, and therefore cannot measure 300's mean abundance and dispersion for the corrected abundances.}
\begin{tabular}{ccc}
\hline
Element & Mean & Dispersion \\
\hline 
$\mbox{[Fe/H]}$ & $-1.39 \pm 0.03 $ & $ < 0.06 $ \\
$\mbox{[C/Fe]}$ & $-0.68 \pm 0.11$ & $ 0.22 _{-0.16} ^{+0.15} $ \\
$\mbox{[N/Fe] }$ & $-0.04 \pm 0.13$ & $ 0.04 _{-0.02} ^{+0.08} $ \\
$\mbox{[O/Fe] }$ & $0.20 \pm 0.05$ & $ 0.03 _{-0.02} ^{+0.05} $ \\
$\mbox{[Na/Fe]}$  & $-0.24 \pm 0.07$ & $ 0.14 _{-0.07} ^{+0.08}$ \\
$\mbox{[Mg/Fe]}$  & $0.08 \pm 0.05$ & $ 0.03 _{-0.02} ^{+0.03} $ \\
$\mbox{[Si/Fe]}$  & $0.25 \pm 0.04$ & $ 0.03 _{-0.02} ^{+0.03}$ \\
$\mbox{[K/Fe]}$  & $0.51 \pm 0.11$ & $ 0.04 _{-0.02} ^{+0.08} $ \\
$\mbox{[Ca/Fe]}$  & $0.15 \pm 0.05 $ & $ 0.03 _{-0.02} ^{+0.03} $ \\
$\mbox{[Ti/Fe]}$  & $0.02 \pm 0.07$ & $ 0.05 _{-0.03} ^{+0.08}$ \\
$\mbox{[Mn/Fe]}$  & $-0.38 \pm 0.06$ & $ 0.03 _{-0.02} ^{+0.04} $ \\
$\mbox{[Ni/Fe]}$  & $-0.11 \pm 0.06$ & $ 0.03 _{-0.02} ^{+0.04} $ \\
$\mbox{[Zn/Fe]}$  & $-0.24 \pm 0.06$ & $ 0.04 _{-0.02} ^{+0.07}$ \\
$\mbox{[Sr/Fe]}$  & $-0.13 \pm 0.14 $ & $ 0.34 _{-0.10} ^{+0.13}$ \\
$\mbox{[Y/Fe] }$ & $-0.47 \pm 0.07$ & $ 0.08 _{-0.06} ^{+0.13}$ \\
$\mbox{[Ba/Fe]}$  & $-0.08 \pm 0.19 $ & $ 0.46 _{-0.11} ^{+0.18} $ \\
$\mbox{[Eu/Fe]}$  & $0.30 \pm 0.06$ & $ 0.12 _{-0.06} ^{+0.07} $ \\
\hline
\end{tabular}
\end{table}

\begin{table}
\caption{\label{tab:enrich}A comparison of the abundances of elements typically enriched or depleted in second-population globular cluster stars. The 1st-Pop Mean abundances are the most probable mean abundances for the first-population stars in 300S, as inferred from five typical stars in our sample (excluding the three anomalous stars). We compare this to the abundance found in our enriched star J1020$+$1555 and give the significance of the star's abundances with respect to the normal star population.}
\begin{tabular}{lrrr}
\hline
Element & 1st-Pop Mean & J1020$+$1555 Abundance & Significance  \\
\hline
$\mbox{[Fe/H]}$       & $-1.39 \pm 0.04$ & $ -1.31 \pm 0.07 $ & $ 0.95 $ \\
$\mbox{[C/Fe]}$       & $-0.74 \pm 0.11$ & $ -0.49 \pm 0.21 $ & $ 1.04 $ \\
$\mbox{[N/Fe]}$       & $-0.10 \pm 0.14$ & $ 0.18 \pm 0.23 $  & $ 1.04 $ \\
$\mbox{[O/Fe]}$       & $0.25  \pm 0.06$ & $ 0.03 \pm 0.17 $  & $ 1.22 $ \\
$\mbox{[Na/Fe]}$      & $-0.35 \pm 0.06$ & $ 0.00 \pm 0.10 $  & $ 2.91 $ \\
$\mbox{[Mg/Fe]}$      & $0.05  \pm 0.06$ & $ 0.07 \pm 0.09 $  & $ 0.18 $ \\
\hline
\end{tabular}
\end{table}

\begin{table}
\caption{\label{tab:binary}A comparison of the abundances of elements typically enriched in CH stars. The 1st-Pop Mean abundances are the most probable mean abundances for the first-population stars in 300S, as inferred from five typical stars in our sample (excluding the three anomalous stars). We compare this to the abundance found in our CH star J1001$+$1554 and give the significance of the star's abundances with respect to the normal star population.}
\begin{tabular}{cccc}
\hline
Element & 1st-Pop Mean & J1001$+$1554 Abundance &Significance  \\
\hline
$\mbox{[Fe/H]}$ & $-1.39 \pm 0.04$ & $-1.38 \pm 0.06$ & 0.16 \\
$\mbox{[C/Fe]}$ & $-0.74 \pm 0.11$ & $-0.09 \pm 0.20$ & 2.80 \\
$\mbox{[Sr/Fe]}$ & $-0.22 \pm 0.08$  & $0.69 \pm 0.13$ & 6.06 \\
$\mbox{[Ba/Fe]}$ & $-0.27 \pm 0.11$  & $0.98 \pm 0.11$ & 8.47 \\
$\mbox{[Eu/Fe]}$ & $0.24 \pm 0.06$  & $0.48 \pm 0.08$ & 2.50 \\
\hline
\end{tabular}
\end{table}

\section{Discussion}
\label{sec:discuss}

\subsection{Globular Cluster, Not a Dwarf Galaxy}
Previous studies of 300S struggled to determine whether the progenitor of the stellar stream was a dwarf galaxy or a globular cluster.
The stream was found by \citet{Fu2018} to have a metallicity dispersion.
\citet{Li2022}, however, measured no metallicity dispersion.
To settle the contradicting results, we re-observed the most metal-poor star in \citet{Fu2018} with MIKE twice to verify its membership.
The star's measured radial velocity is inconsistent with the stream moving at 300 \kms;
further, the star shows significant variations in its radial velocities over time: on the two occasions the star was observed with MIKE, the observed radial velocity was measured to be ${\sim}260$ \kms.
\sfive measured a velocity of ${\sim}275$ \kms.
Lastly, the SEGUE data used by \citet{Fu2018} indicates a velocity of ${\sim}297$ \kms.
These large variations in radial velocity (>50 \kms) measured over several years imply the star is a part of a binary system. The average velocity is not similar enough to the stellar stream to be considered a member, as can be observed in Figure~\ref{fig:sky_map}.
Excluding this star from the analysis, the 300S stars have an insignificant metallicity spread.

\citet{Frebel2013} also suggested the progenitor of 300S was a dwarf galaxy due to its low magnesium.
We confirm the low [Mg/Fe] abundance, though we disagree with the interpretation.
The GCs that formed \textit{in situ} essentially sample the [Mg/Fe] ratio at different ages of the Milky Way.
Low [Mg/Fe] abundance can simply indicate that a stellar system formed \textit{ex situ}, such as the low Mg abundance of the S2 stream which may be a part of the Helmi stream \citep{Aguado2021b}.
The progenitor of 300S likely formed \textit{ex situ}.
The stream's trajectory is most similar to that of the GSE and its associated remnant GCs \citep{Myeong2018, Massari2019, Forbes2020, Callingham2022, Limberg2022}.
\citealt{Fu2018} found an orbit similar to the Virgo Overdensity, which has since been determined to be debris from the GSE merger, e.g., \citealt{Simon2019, Iorio2019, Naidu2021, Perottoni2022}.
Similarly, \citet{Li2022} calls attention to 300S's high eccentricity and its position close to GSE in the $E_\mathrm{tot}-L_Z$ plane to further suggest their association.
These matching orbits indicate that it could have formed in GSE prior to the latter's merger with the Milky Way.

\begin{figure*}
    \centering
    \includegraphics[width=0.9\textwidth]{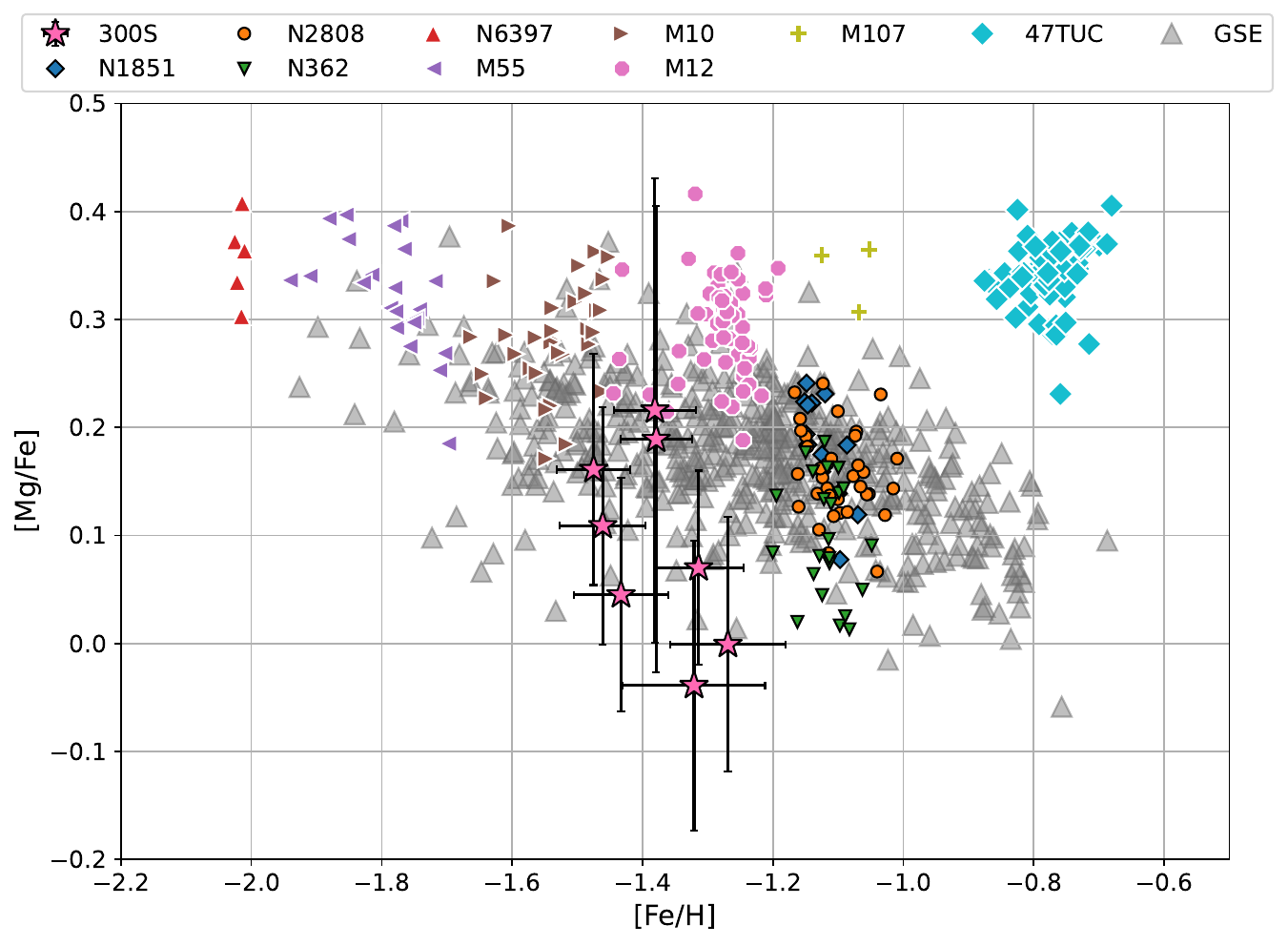}
    \caption{\label{fig:mg_comp_v3}Magnesium to iron ratio in stars of Milky Way GCs, the stellar stream 300S, the GSE and GSE GCs.
    The Milky Way and GSE GCs star abundances are colored points with white and black outlines respectively.
    The abundances of GSE stars are represented by gray triangles, while our 300S measurements are represented by pink stars with black errorbars.
    GCs sample the [Mg/Fe] ratio of its host galaxy at its time of formation.
    The 1P stars for each cluster are the most metal rich at the top of the plot, while 2P stars have depleted Mg and stretch to the bottom of the figure.
    300S does not resemble Milky Way GCs, but matches the GSE, indicating it formed \textit{ex situ}.
   This chemical resemblance further supports the stream’s dynamic association with GES given its highly eccentric and retrograde orbit: its energy and angular momenta similarly match that of GES \citep{Fu2018}.
    The globular cluster abundances here are data APOGEE data from \citet{Meszaros2020}, while the GSE abundances are APOGEE data from \citet{Limberg2021}.}
\end{figure*}

Indeed, the [Mg/Fe] observed in 300S matches the abundances found in GSE.
In Fig.~\ref{fig:mg_comp_v3}, we compare the magnesium abundances of stars in 300S (represented by pink stars with black errorbars) to abundances in other systems.
The gray semi-transparent triangles and symbols with dark outlines are APOGEE abundances for GSE stars and GSE GC stars, as identified by \citet{Limberg2021}.
The symbols with white outlines are APOGEE abundances for Milky Way GC stars identified by \citet{Meszaros2020}, though we adopt the APOGEE DR17 abundances due to a significant zero-point offset. 
The most Mg-rich stars from each cluster represent the 1P stars for the respective system, as enriched stars in MSPs become depleted of Mg.
We define the 1P stars using a cut in [Al/Fe]; since enriched stars exhibit high aluminum abundances, we make a cut at [Al/Fe] $< 0.25$ to determine which stars are unenriched.
For the Milky Way GC stars, the resulting Mg-rich 1P stars hover around [Mg/Fe] = 0.35.
The most Mg-rich stars of 300S fall alongside the Mg abundances of GSE and GSE cluster stars with an average [Mg/Fe] $\approx$ 0.2 \citep{Belokurov2023}.
300S abundances more closely aligns with the Mg content found in GSE and its clusters, further indicating along with the similar orbital properties that they may have a common origin, or that the former formed inside of the latter.

\begin{figure*}[h]
    \centering
    \includegraphics[width=0.9\textwidth]{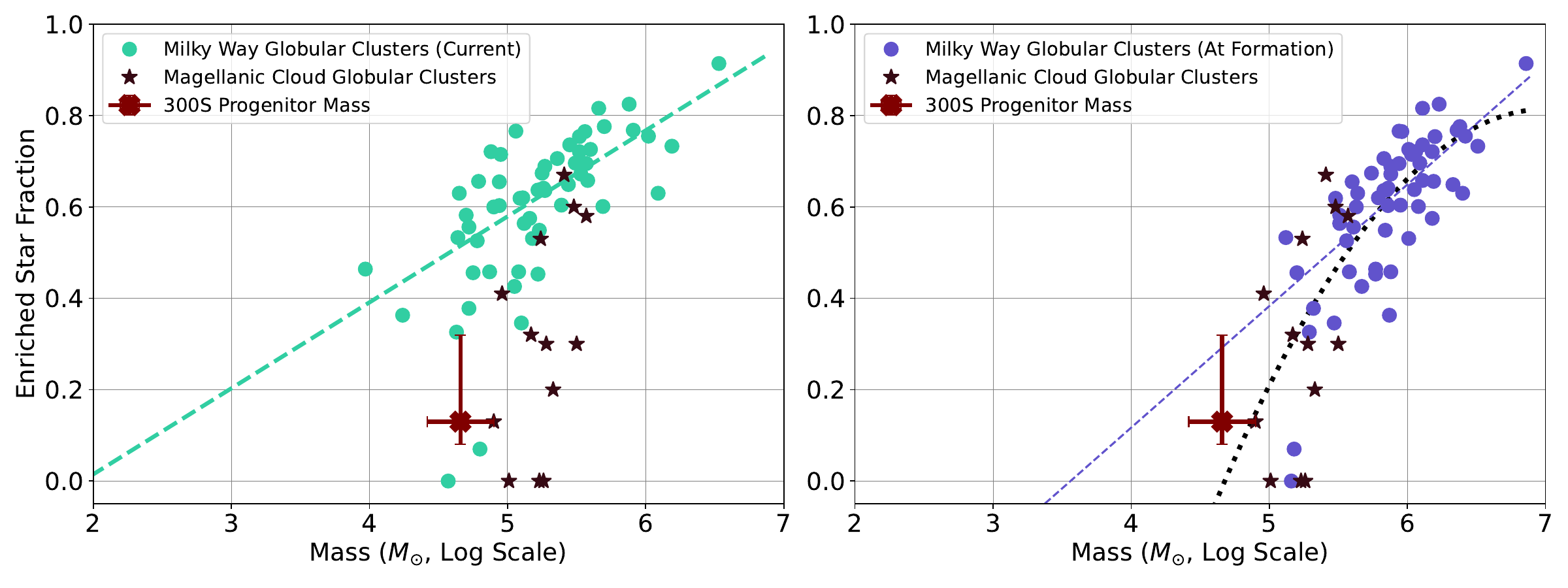}
    \caption{\label{fig:enrich}The current masses (green circles, left) and initial masses (purple circles, right) of intact Milky Way GCs \citep{Milone2017, Baumgardt2019}. The measurements for the Magellanic Clouds come from \citet{Mucciarelli2009, Mateluna2012, Mucciarelli2014, Hollyhead2017, Niederhofer2017a, Niederhofer2017b, Martocchia2017, Hollyhead2018, Zhang2018, Martocchia2018}, and \citet{Hollyhead2019} and were compiled by \citealt{Gratton2019}. The lower limit for 300S's mass is its current stream mass, while the upper limit of its mass is calculated from based on a fully dissolved cluster as described in Section~\ref{sec:mass}. In the left panel, we see a correlation between the present-day mass of intact Milky Way GCs and the fraction of enriched stars found therein. If we extrapolate a linear relationship, we would expect a system with a mass of $10^{4.5} \msun$ like 300S would have about half of its stars enriched. Instead, just one of our eight stars exhibited enrichment. In the right panel, we compare the initial masses and the fraction of enriched stars for these same systems. We find a much tighter correlation. If we again assume a linear relationship, 300S still appears underenriched for its mass, but does not appear as significantly displaced. A quadratic relation, however, may be better at representing a critical threshold limit needed to create stellar enrichment. Extrapolating from this relationship (the black, dotted line), 300S appears more closely in line with predictions. In both panels, 300S is marginally more enriched than two other Milky Way globular clusters: Ruprecht 106 and Terzan 8.}
\end{figure*}

\subsection{Multiple Populations and Initial Mass}
We further confirm that the progenitor of 300S was a globular cluster by detecting a second-population star, J1020$+$1555.
However, the stream does not contain as many enriched stars as we had initially expected.
A correlation exists between the mass of intact Milky Way GCs and the fraction of enriched stars found therein, as seen in the left panel of Fig.~\ref{fig:enrich}.
Assuming a binomial distribution, we infer that for just one of eight stars to be enriched, the stream is likely to have an inherent enrichment fraction of f$_{\text{enrich}} = 0.13^{+0.19}_{-0.05}$, as is reflected in the figure.
If we choose to exclude the cool star entirely from the sample due to its ambiguous enrichment, we calculate the stream's inherent enrichment fraction to be f$_{\text{enrich}} = 0.14^{+0.21}_{-0.06}$.
We estimated 300S's current stream mass to be 10$^{4.5}$\msun, as described in Section ~\ref{sec:mass}.
The total stream mass should be higher than this, since more stream members beyond the footprint of \sfive~were not included in this estimate. 
Interpolating a linear relationship between the current mass of intact Milky Way GCs and their fraction of enriched stars (as seen in the left panel of Fig.~\ref{fig:enrich}), we expected half of the stars in our sample to be enriched.
Given our small sample size, we cannot completely exclude a 50\% enrichment fraction, but we proceed taking the observed enriched fraction at face value.

With just one of the eight stars being enriched, 300S appears  underenriched when compared to the present day masses of GCs, but is fairly reasonable when we compare the \textit{progenitor} mass to the \textit{initial} masses of intact Milky Way GCs.
Intact Milky Way GCs have lost, on average, 80\% of their initial mass over their lifetimes \citep{Baumgardt2019} due to a combination of evaporation and tidal stripping.
In Section~\ref{sec:mass} we estimated limits on the progenitor mass of 300S to be in the range of 10$^{4.5-4.9}$\msun.
In the right panel of Fig.~\ref{fig:enrich}, we compare the initial masses of intact Milky Way GCs with our best constraints on the initial mass of 300S.
300S does not appear anomalous in this framework, as it appears relatively in line with the relationship of enrichment fraction to initial mass as can be seen in other clusters containing MSPs \citep{Belokurov2023}.
The stream progenitor could still be considered underenriched if we assume a linear relationship between fractional enrichment and initial mass of the globular cluster, but it could also indicate a non-linear relationship between the cluster parameters.

The masses of GCs in the Large and Small Magellanic Clouds (represented by black stars in Fig.~\ref{fig:enrich}, as measured by \citealt{Mucciarelli2009, Mateluna2012, Mucciarelli2014, Hollyhead2017, Niederhofer2017a, Niederhofer2017b, Martocchia2017, Hollyhead2018, Zhang2018, Martocchia2018}, and \citealt{Hollyhead2019} and compiled by \citealt{Gratton2019}) support this theory.
It has been suggested that these clusters may have lost less mass, since the Magellanic Clouds are significantly less massive than the Milky Way \citep[e.g.,][]{Milone2020}.
We therefore assume that the Magellanic Cloud GC initial masses are relatively similar to the masses we see now.
These clusters' masses fall in line with calculated initial masses of intact Milky Way GCs, and trend towards having a non-linear increase in enrichment above 10$^5$\msun.

It has been suggested a minimum critical mass must be achieved in order to create MSPs \citep[e.g.,][]{Gratton2019}, though this threshold mass is currently undetermined.
The initial mass of 300S may hover near the critical threshold, allowing the unusually low enrichment we see here.
Further improvements to modelling of GCs to better calculate initial masses will allow us to more strongly constrain the mass-enrichment relationship.
Additionally, spectroscopic follow-up of more globular cluster stellar streams may further serve as a useful probe to finding a critical mass needed to achieve chemical enrichment found in most GCs.
Theoretically, \citet{BlandHawthorn2010} suggests that a single-population globular cluster is achievable up to an initial mass of 10$^{6.5}$ \msun. If this corresponds to the minimum mass threshold, it would suggest significant revisions are needed to globular cluster mass loss.

\subsection{CH Star and Cluster Dynamics}
We find a CH star in our stellar stream, enriched in carbon, barium, yttrium, and strontium.
In this text, we have referred to our \textit{s}-process enhanced star as a CH star, though it also fits the criteria of a Ba star (but is too metal-rich to be considered a CEMP-s star).
We therefore look to previous literature on Ba stars to shed light on our \textit{s}-process enhanced star, specifically to understand their rarity.

Binary systems are scarce in globular clusters compared to the field, perhaps due to their high density disrupting bound binary orbits \citep{Cote1997}.
Since Ba stars are thought to form due to accretion from a companion star \citep{McClure1980}, we would expect Ba stars to be similarly unusual in globular clusters.

For comparison, \citet{Luck1991} found the fraction of Ba stars in the field to be $\sim$2\%.
The more recent study \citet{Norfolk2019} used the LAMOST survey to find that approximately 0.2\% of their sample of 454,180 giant stars to be s-process enhanced. 
26\% of these stars (0.05\% of the total sample) have detectable amounts of Ba, though this may be an underestimate due to limitations from LAMOST's low-resolution spectra.
Therefore the rate of Ba stars in the field is probably in the range 0.05\% to 2\% of stars.

\citet{DOrazi2010} specifically explored the rate of Ba stars in globular clusters and found 5 of the 1,205 red giant stars observed (0.4\%) were Ba stars, endorsing previous theories that Ba stars are rare in globular clusters.
However, this rate appeared dependent on whether the star was part of the enriched or unenriched star populations.
The authors found four of the five Ba stars were unenriched stars, estimating that about $\sim$2\% of unenriched stars in globular clusters.
The Ba star rate in enriched star populations can therefore be calculated to be $\sim$0.1\% using Bayes' theorem.

Finding a Ba star in our small star sample is therefore very surprising and unexpected.
Assuming a binomial distribution, we can infer that the intrinsic rate of Ba stars in 300S's unenriched star population is f$_{\text{Ba}} = 14^{+21}_{~-5}$\%. 
This is significantly higher than the rate of $\sim$2\% found by \citet{DOrazi2010} for unenriched star populations in intact star clusters.
This could indicate that the disruption of globular clusters allows for binary systems to survive longer than would've been possible in an intact cluster environment, enabling them to reach a stage where accretion from a binary companion can be achieved.
Further spectroscopic study of globular cluster stellar streams could help clarify whether streams in general provide a fruitful environment for the survival of binaries.

\section{Conclusion}
\label{sec:conclusion}

The discovery and analysis of 300S has led to interesting implications for the formation of MSPs and of the survival of binaries in GCs.
Through our observation of 10 red giant branch candidate stars and subsequent spectroscopic analysis of 8 member stars, we find that:

\begin{itemize}
    \item The progenitor of 300S is a globular cluster. 
    \begin{itemize}
        \item The stream is relatively metal-rich and magnesium-poor when compared to Milky Way GCs, but its orbit indicates that it formed \textit{ex situ,} perhaps inside the GSE. 
        The stream's metallicity and Mg abundance is a reflection of its host galaxy, not the nature of its progenitor, as can be seen in Fig.~\ref{fig:mg_comp_v3}.
        \item An analysis by \citet{Fu2018} indicated 300S had a measurable metallicity dispersion, indicating its progenitor was a dwarf galaxy. However, we found the most metal-poor star in the sample to have a discrepant radial velocity (shown in Table~\ref{tab:obs}).
        The remaining stars have no significant spread in metallicity.
        \item We identify one star, J1020$+$1555, as a second-population star, enriched in sodium and nitrogen while depleted in oxygen.
        MSPs are a hallmark of GCs, confirming the nature of the stream's progenitor.
        The comparison of these key elements are shown in Fig.~\ref{fig:second_pop} and Fig.~\ref{fig:second_pop_corr}.
    \end{itemize}
    \item The progenitor of 300S straddled the threshold of the critical mass for making MSPs. 
    We calculate the initial mass of 300S based on its orbital parameters and compare it to the initial masses of intact GCs as calculated by \citet{Baumgardt2019}.
    We find it to have an initial mass of M$_{\text{Ini}}$ = 10$^{4.5-4.9}$ \msun, which aligns with the initial masses of GCs with small fractions of enriched stars.
    This comparison is shown in Fig.~\ref{fig:enrich}.
    300S is therefore a good benchmark for the critical mass threshold of MSPs.
    \item We find a CH star, which is unusual inside of intact GCs.
    The star is shown to be enriched in carbon and \textit{s}-process elements such as strontium, yttrium, and barium, as shown in Fig.~\ref{fig:binary}.
    This could be an indication that stellar streams may provide a more suitable environment for the formation and sustained lifetime of binary systems than intact GCs.
\end{itemize}

Future work is needed to further explore the relationship between initial mass and MSPs in GCs.
Exploring globular cluster streams would allow us to further probe this relationship using \textit{ex situ} sources.
Another way to confirm this would be to look at a galaxy of similar size, such as M31, and perform a similar calculation as found in \cite{Baumgardt2019}.
This would allow us to test for a consistency in the relationship between mass and fraction of enriched stars in galaxies outside of our own.
By better constraining the relationship between initial mass and enrichment, we could then reverse engineer the correlation to estimate the mass loss of GCs in a galaxy.
The trend of GCs' fractional enrichment compared to their mass could then be compared to the confirmed initial mass threshold for multiple population formation.
This be a novel way to estimate the initial mass for GCs in distant galaxies.

\section*{Acknowledgements}

This paper includes data gathered with the 6.5~meter Magellan Telescopes located at Las Campanas Observatory, Chile.
We thank all the LCO staff for enabling astronomical observations during a global pandemic.
This paper includes data obtained with the Anglo-Australian Telescope in Australia. We acknowledge the traditional owners of the land on which the AAT stands, the Gamilaraay people, and pay our respects to elders past and present.
For the purpose of open access, the author has applied a Creative Commons Attribution (CC BY) licence to any Author Accepted Manuscript version arising from this submission.

We thank Jeffrey Gerber for his insightful feedback about CH stars.
S.A.U. and A.P.J. acknowledge financial support from NSF grant AST-2206264.
T.S.L. acknowledges financial support from Natural Sciences and Engineering Research Council of Canada (NSERC) through grant RGPIN-2022-04794.
A.B.P. is supported by NSF grant AST-1813881. 

This work has made use of data from the European Space Agency (ESA) mission
{\it Gaia} (\url{https://www.cosmos.esa.int/gaia}), processed by the {\it Gaia}
Data Processing and Analysis Consortium (DPAC,
\url{https://www.cosmos.esa.int/web/gaia/dpac/consortium}). Funding for the DPAC
has been provided by national institutions, in particular the institutions
participating in the {\it Gaia} Multilateral Agreement.

This research has made use of the SIMBAD database, operated at CDS, Strasbourg, France \citep{Simbad}.
This research has made use of NASA’s Astrophysics Data System Bibliographic Services.

\section*{Data Availability}

The individual line measurements are provided as supplementary material. The reduced spectra can be obtained by reasonable request to the corresponding author.



\bibliography{main}
\bibliographystyle{mnras}

\bsp	

\begin{table*}
\caption{The following is a sample table of data for the star 0956+1603. Full data tables are available online.}
\begin{tabular}{llccrrrrrrrrrr}
Star & El. & $N$ & ul & $\log \epsilon$ & [X/H] & $\sigma_{\text{[X/H]}}$ & [X/Fe] & $\sigma_{\text{[X/Fe]}}$ & $\Delta_T$ & $\Delta_g$ & $\Delta_v$ & $\Delta_M$ & $s_X$ \\
\hline
0956+1603      & C-H   &   2 &     &$+6.51$&$-1.95$&  0.23 &$-0.63$&  0.22 &  0.18 & -0.01 & -0.01 &  0.21 &  0.04 \\
0956+1603      & C-N   &   1 &     &$+6.43$&$-1.40$&  0.37 &$-0.08$&  0.34 &  0.34 &  0.01 & -0.02 &  0.27 &  0.00 \\
0956+1603      & O I   &   1 &     &$+7.65$&$-1.04$&  0.16 &$+0.28$&  0.16 &  0.03 &  0.10 &  0.01 &  0.09 &  0.00 \\
0956+1603      & Na I  &   2 &     &$+4.49$&$-1.76$&  0.12 &$-0.44$&  0.12 &  0.10 &  0.00 & -0.01 &  0.00 &  0.00 \\
0956+1603      & Mg I  &   4 &     &$+6.27$&$-1.36$&  0.10 &$-0.04$&  0.13 &  0.06 & -0.05 & -0.03 &  0.02 &  0.09 \\
0956+1603      & Si I  &   4 &     &$+6.37$&$-1.15$&  0.10 &$+0.17$&  0.12 &  0.07 &  0.03 & -0.01 &  0.02 &  0.00 \\
0956+1603      & K I   &   2 &     &$+4.42$&$-0.65$&  0.28 &$+0.67$&  0.25 &  0.20 & -0.05 & -0.13 & -0.04 &  0.00 \\
0956+1603      & Ca I  &   7 &     &$+5.18$&$-1.12$&  0.12 &$+0.20$&  0.14 &  0.09 &  0.03 &  0.05 &  0.01 &  0.05 \\
0956+1603      & Ti I  &   2 &     &$+4.05$&$-0.85$&  0.33 &$+0.47$&  0.32 &  0.17 &  0.00 & -0.00 &  0.00 &  0.34 \\
0956+1603      & Mn I  &   6 &     &$+3.66$&$-1.69$&  0.20 &$-0.37$&  0.19 &  0.15 & -0.00 & -0.01 &  0.00 &  0.15 \\
0956+1603      & Fe I  &  20 &     &$+6.09$&$-1.32$&  0.11 &$+0.00$&  0.00 &  0.07 &  0.01 &  0.04 &  0.02 &  0.15 \\
0956+1603      & Fe II &   5 &     &$+6.17$&$-1.34$&  0.12 &$+0.00$&  0.00 & -0.03 &  0.10 & -0.02 &  0.06 &  0.03 \\
0956+1603      & Ni I  &   4 &     &$+4.58$&$-1.49$&  0.23 &$-0.17$&  0.22 &  0.14 &  0.03 &  0.03 &  0.02 &  0.23 \\
0956+1603      & Zn I  &   2 &     &$+2.98$&$-1.57$&  0.15 &$-0.25$&  0.15 &  0.03 &  0.07 & -0.06 &  0.05 &  0.07 \\
0956+1603      & Sr II &   1 &     &$+1.12$&$-1.75$&  0.38 &$-0.41$&  0.28 &  0.21 &  0.06 & -0.11 &  0.20 &  0.00 \\
0956+1603      & Y II  &   4 &     &$+0.39$&$-1.80$&  0.19 &$-0.46$&  0.12 &  0.05 &  0.10 & -0.02 &  0.08 &  0.04 \\
0956+1603      & Ba II &   5 &     &$+0.68$&$-1.50$&  0.27 &$-0.16$&  0.20 &  0.04 &  0.05 & -0.13 &  0.06 &  0.26 \\
0956+1603      & Eu II &   1 &     &$-0.68$&$-1.20$&  0.17 &$+0.14$&  0.16 & -0.01 &  0.10 &  0.02 &  0.08 &  0.00 \\
\hline 
\end{tabular}
\end{table*}

\end{document}